\documentclass[11pt]{article}
\usepackage[utf8]{inputenc}
\usepackage[a4paper, total={6in, 8in}]{geometry}
\usepackage{graphicx}
\usepackage{bm}
\usepackage{dsfont}  
\usepackage[misc]{ifsym}
\usepackage{amsfonts,amssymb}
\usepackage{hyperref}
\hypersetup{colorlinks=true, linkcolor=blue,
anchorcolor=blue, citecolor=blue}
\usepackage[round]{natbib}
\usepackage{amsmath}
\usepackage{bbm}
\usepackage{authblk}
\usepackage{setspace}
\usepackage{tabularx}
\usepackage{indentfirst}
\usepackage{siunitx}
\usepackage{booktabs} 
\usepackage{multirow}
\usepackage{adjustbox}
\usepackage{makecell}
\usepackage{threeparttable}
\usepackage{appendix}
\usepackage{float}
\usepackage{pdfpages}
\usepackage[linesnumbered,ruled,vlined]{algorithm2e}
\DontPrintSemicolon
\usepackage{algpseudocode}

\SetKwInput{KwInput}{Input}               
\SetKwInput{KwOutput}{Output}
\SetKwInput{KwProcedure}{Procedure}

\title{Gradient boosted multi-population mortality modelling \\with high-frequency data}
\author[1]{Ziting Miao}
\author[1]{Han Li}
\author[1]{Yuyu Chen}
\affil[1]{Department of Economics, University of Melbourne, Australia}

\date{}

\begin{document}
\maketitle

\begin{abstract}

High-frequency mortality data have attracted growing attention, but their use has largely been confined to specific applications rather than general modelling and forecasting. Such data pose new challenges to traditional mortality models due to pronounced seasonal patterns and short-term fluctuations. To address these challenges and produce more accurate forecasts with the high-frequency mortality data, this paper introduces a novel integration of gradient boosting techniques into traditional stochastic mortality models under a multi-population setting. Our key innovation lies in using the Li and Lee model as the weak learner within the gradient boosting framework, replacing conventional decision trees. Empirical studies are conducted using weekly mortality data from 30 countries (Human Mortality Database, 2015--2019). Empirical evidence highlights that the proposed methodology not only enhances model fit by accurately capturing underlying mortality trends and seasonal patterns, but also achieves superior forecast accuracy, compared to the benchmark models. We also investigate a key challenge in multi-population mortality modelling: how to select appropriate sub-populations with sufficiently similar mortality experiences. A comprehensive clustering exercise is conducted based on mortality improvement rates and seasonal strength. The empirical results demonstrate that our proposed model maintains strong forecast accuracy across different clustering configurations, thereby reducing the need for extensive data preprocessing.

\end{abstract}

\clearpage 
\newpage

\section{Introduction}
Most existing research on mortality modelling uses annual data, where short-term mortality fluctuations are often overlooked. Nevertheless, modelling high-frequency data is increasingly important as it allows for capturing short-term mortality dynamics, seasonal fluctuations, and rapid responses to shocks such as epidemics or extreme weather events. While some recent studies have utilised monthly data to improve timely forecasts and support public health planning \citep[see e.g.,][]{leger2024month, li2024hierarchical}, the modelling of weekly mortality data remains relatively under-researched and has largely focused on the analysis of excess mortality using the Short-Term Mortality Fluctuations (STMF) data \citep{stmf2021short}. This strand of literature highlights substantial discrepancies between excess deaths and officially reported COVID-19 fatalities, as well as the importance of accounting for temporal dynamics, demographic heterogeneity, and cross-country dependencies in mortality modelling and forecasting \citep[see e.g.,][]{kontis2020magnitude, jdanov2021short, karlinsky2021world, vanella2021assessing, hanebeck2025dependence}. More importantly, there is an urgent need for a new mortality modelling framework specifically designed for high-frequency mortality analysis - highlighting a crucial gap in current research.

The COVID-19 pandemic has accelerated the availability and use of weekly mortality data, which is originally collected to assess the effectiveness of national containment strategies. It enables timely and precise monitoring of recent mortality dynamics, with minimal reporting delays \citep{nepomuceno2022sensitivity}. However, weekly mortality data introduce added complexity, including stronger seasonal patterns and short-term fluctuations that challenge traditional stochastic mortality models, which are designed for lower-frequency data. To address this, we propose a modelling framework specifically designed for high-frequency settings by integrating multi-population stochastic mortality models with gradient boosting techniques. This approach uncovers seasonal patterns in both shared and country-specific mortality trends, while incorporating a highly flexible structure that accommodates heterogeneous mortality experiences across countries and age groups.

The literature on mortality modelling includes numerous efforts to develop stochastic models, both for single-population contexts \citep[see e.g.,][]{osmond1985using, lee1992modeling, jacobsen2002long, renshaw2006cohort, hyndman2007robust} and for multi-population settings \citep[see e.g.,][]{li2005coherent, russolillo2011extending, hyndman2013coherent, danesi2015forecasting, lam2023multipopulation, dimai2024multi}. In this paper, we focus on multi-population mortality models for the following reasons. First, weather-related events such as heatwaves, cold spells, and regional storms typically span across national borders, affecting multiple neighbouring countries simultaneously. Accordingly, multi-population modelling allows for the effective capture of joint mortality dynamics, offering valuable insight into the influence of common shocks across interconnected populations. Second, country-specific weekly mortality rates often exhibit considerable noise and variability, which may obscure underlying trends \citep{arik2025measuring}. This issue can be addressed by extracting the common trends across populations, which helps reveal smoother seasonal patterns by ``borrowing" experience across countries, thereby generating more coherent future projections. Models based on weekly mortality data require a greater level of flexibility than traditional stochastic mortality models can provide, as they must simultaneously capture age-specific seasonal patterns (e.g., the elderly are more sensitive to temperature extremes), and geographically different seasonal trends (e.g., opposing Northern and Southern Hemisphere cycles).

Gradient boosting has emerged as a powerful tool in mortality modelling due to its flexibility and superior ability to uncover hidden patterns in the data. \cite{deprez2017machine} incorporate a wide range of individual-level features, including gender, age, year, and cohort to better assess the impact of contributing factors on mortality. Additionally, they demonstrate how regression tree boosting can help detect more patterns and facilitate cause-of-death estimates. Tree-based ensemble methods are also shown to improve both model fitting and forecast performance \citep[see e.g.,][]{levantesi2019application, bjerre2022tree, qiao2024machine}. Traditional gradient boosting algorithms iteratively fit weak learners to residuals, gradually improving model performance until a stopping criterion is met. Common implementations include the Gradient Boosting Machine \citep{friedman2001greedy}, XGBoost \citep{chen2016xgboost}, and LightGBM \citep{ke2017lightgbm}.

The integration of stochastic mortality models as weak learners within a gradient boosting framework is first explored by \cite{li2025boosting}. They propose a gradient boosted Lee--Carter model for annual mortality data, where long-term trends dominate and seasonal effects are negligible. To forecast US state-level mortality rates, their approach incorporates spatial dependence via a penalty constraint that shrinks mortality rates according to a predefined adjacency-based matrix reflecting geographic proximity. However, this dependence structure is not applicable to general multi-population settings, particularly those involving countries across different hemispheres.

Instead of the shrinkage approach, we propose a multi-population gradient boosting framework with the Li and Lee model \citep{li2005coherent} being the weak learner, which we refer to as the Gradient Boosted Li and Lee model, specifically designed to accommodate high-frequency mortality data. Through iterative applications of the Li and Lee model, the proposed method can capture multiple layers of trends and seasonal variations across all age groups and countries where these patterns are often under-represented in a single fitting. Final forecasts are obtained by aggregating the sequence of fitted Li and Lee models, weighted by an optimally tuned learning rate. Moreover, our proposed algorithm terminates once all underlying patterns have been captured, as indicated by residuals approaching white-noise behaviour, or when the maximum in-sample fitting error across all horizons, countries, and age groups falls below a specific threshold. A predefined maximum number of iterations is also imposed to prevent potential overfitting. This addresses a key limitation of the gradient boosted Lee--Carter model \citep{li2025boosting}, which determines its stopping criterion solely based on convergence between consecutive fits. Their approach may not be suitable for high-frequency data, where residual differences diminish rapidly between iterations. In contrast, our proposed framework employs a more comprehensive training procedure that preserves informative features across iterations, enhancing the capacity of our proposed model to extract meaningful trends and seasonal signals from high-frequency mortality data, which are often missed by simpler models (see Section \ref{sec:fitting} for more details).

In multi-population mortality modelling, the grouping strategy for sub-populations plays an important role in the identification of the shared mortality trends. \cite{hyndman2013coherent} suggest grouping sub-populations based on predefined characteristics such as sex or states within a country. Departing from this conventional approach, we employ data-driven clustering techniques to identify groups of countries that exhibit similar mortality dynamics, irrespective of economic or geographic classifications. While previous studies primarily group countries based on age-specific mortality patterns \citep[see e.g.,][]{hatzopoulos2013common, andreopoulos2022different, boonen2026low}, or form groups using a combination of country and age characteristics \citep[see e.g.,][]{leger2021can, debon2024multipopulation}, we propose to leverage seasonal trends and cyclical structures as the basis for grouping, reflecting the unique properties of high-frequency data. In particular, we highlight the contribution of using seasonal strength, which captures the magnitude of cyclical mortality patterns across populations. In addition to seasonal strength, we consider alternative clustering approaches based on trend slopes and combined features. This provides a broader framework to examine whether the proposed model exhibits improved predictive performance across different data inputs, offering insight into its stable performance under varying configurations.

We apply the proposed gradient boosted Li and Lee model to weekly mortality data from 30 countries, using the Short-Term Mortality Fluctuation series published by the Human Mortality Database for the period 2015 to 2019. To ensure comparability in the timing of mortality peaks and troughs, mortality rates for Southern Hemisphere countries (Australia and New Zealand) are inverted to align with the seasonal patterns of Northern Hemisphere countries. While a random walk with drift is typically used to forecast the time trend under annual data \citep[see e.g.,][]{lee1992modeling, girosi2007understanding, shapovalov2019bayesian}, we incorporate additional Fourier regressors \citep{serfling1963methods} to account for seasonal cyclical patterns, on top of the time series models. The use of sine and cosine terms allows for the joint modelling of short-term seasonal variation and long-term decline in weekly mortality rates. Empirical results demonstrate that the proposed framework outperforms existing benchmark models in terms of both model fit and forecast accuracy. Moreover, the performance of the proposed model is not sensitive to the clustering selection criteria, which reduces the need for extensive data preprocessing or manual adjustments.

The contribution of this paper can be summarised as follows:
\begin{itemize}
\item We are among the first to model and forecast high-frequency mortality rates by introducing a flexible gradient boosting framework that employs the Li and Lee model as the weak learner within a multi-population context. To harmonise seasonal patterns across hemispheres, we propose a reciprocal transformation to observed mortality rates in Southern Hemisphere countries. This framework not only captures the underlying short-term seasonal cycles and declining trends, but also enhances forecast accuracy.

\item We conduct a clustering-based experiment to explore whether certain strategies for grouping countries yield improved performance and to examine the sensitivity of our proposed model to the input data. While certain clustering methods lead to minor improvements and others result in slight performance declines, empirical evidence suggests that our proposed model outperforms the baseline models across all clustering methods. This insensitivity of our framework is further demonstrated by its superior forecast performance, with only marginal differences observed across the three clustering approaches.
\end{itemize}

The remainder of this paper is structured as follows. Section \ref{sec:Data} describes and visualises weekly mortality rates for 30 countries. Section \ref{sec:Method} introduces the proposed gradient boosted Li and Lee model. Section \ref{sec:results} presents and compares the fitting and forecast performance of the proposed model with baseline models. Section \ref{sec:clustering} examines three clustering techniques to group countries with similar mortality experiences and evaluates their forecasting results. Finally, Section \ref{sec:conclusion} concludes the paper and outlines directions for future research. The data and code are available in a GitHub repository at \url{https://github.com/amymiao1019/GBLL}. This paper is accompanied by a Supplementary Material with additional analysis regarding alternative parameter choices of the proposed model.

\section{Weekly mortality data}
\label{sec:Data}

\subsection{Background and description}
In response to the COVID-19 pandemic, the Human Mortality Database launched the Short-Term Mortality Fluctuations (STMF) series \citep{stmf2021short} in May 2020, attracting growing interest in the analysis of excess mortality. \cite{jdanov2021short} develop an online tool to visualise weekly excess deaths across countries, using 2010--2019 as the reference period. Based on STMF data from 2016--2020, \cite{islam2021excess} find that excess deaths in 29 high-income countries substantially exceed officially reported COVID-19 deaths, particularly among children under 15. This result is supported by \cite{karlinsky2021world}, who underscore the need for timely and reliable mortality data to improve forecast accuracy. \cite{kontis2020magnitude} focus on countries with populations over four million and incorporate seasonal patterns, temperature effects, long-term trends, and weekly dependencies into their model. Their findings reveal that individuals aged 65 and older account for 94\% of total excess deaths, both in absolute and relative terms. \cite{vanella2021assessing} apply the Principal Component Analysis to mortality data from 19 European countries. They demonstrate that country-specific models often overlook shared common trends, reducing forecast accuracy. Their study underscores the importance of accounting for mortality correlations across countries, particularly given the transnational spread of pathogens. They also identify substantial cross-country variation in excess mortality, driven by differences in public health responses and demographic structures. Similarly, \cite{arik2025measuring} present a cross-country analysis of excess mortality during 2020--2023, showing broadly similar patterns across measures while highlighting that methodological choices can affect the magnitude and variability of the estimates.

While much of the literature has focused on explaining pandemic-related excess mortality, this paper aims to address the broader need for long-term mortality forecasts using high-frequency data. In this paper, we collect mortality data of 30 countries from the STMF series, excluding those with populations under two million due to the frequent occurrence of zero weekly deaths. Such data sparsity introduces substantial uncertainty and can result in unstable model estimates. We extract national-level mortality rates spanning from the second week of 2015 to the final week of 2019, comprising exactly 52 weeks per year. To address the issue of zero mortality values which result in undefined logarithmic transformations, we choose to exclude the youngest age group (0--14 years old) from our analysis. Consequently, we focus on the remaining four age groups: 15--64, 65--74, 75--84, and 85+ years old.

\subsection{Visualisation}
\label{sec:data_visualisation}
\begin{figure}[h!]
\centering
\includegraphics[width=1\textwidth, height=0.32\textheight]{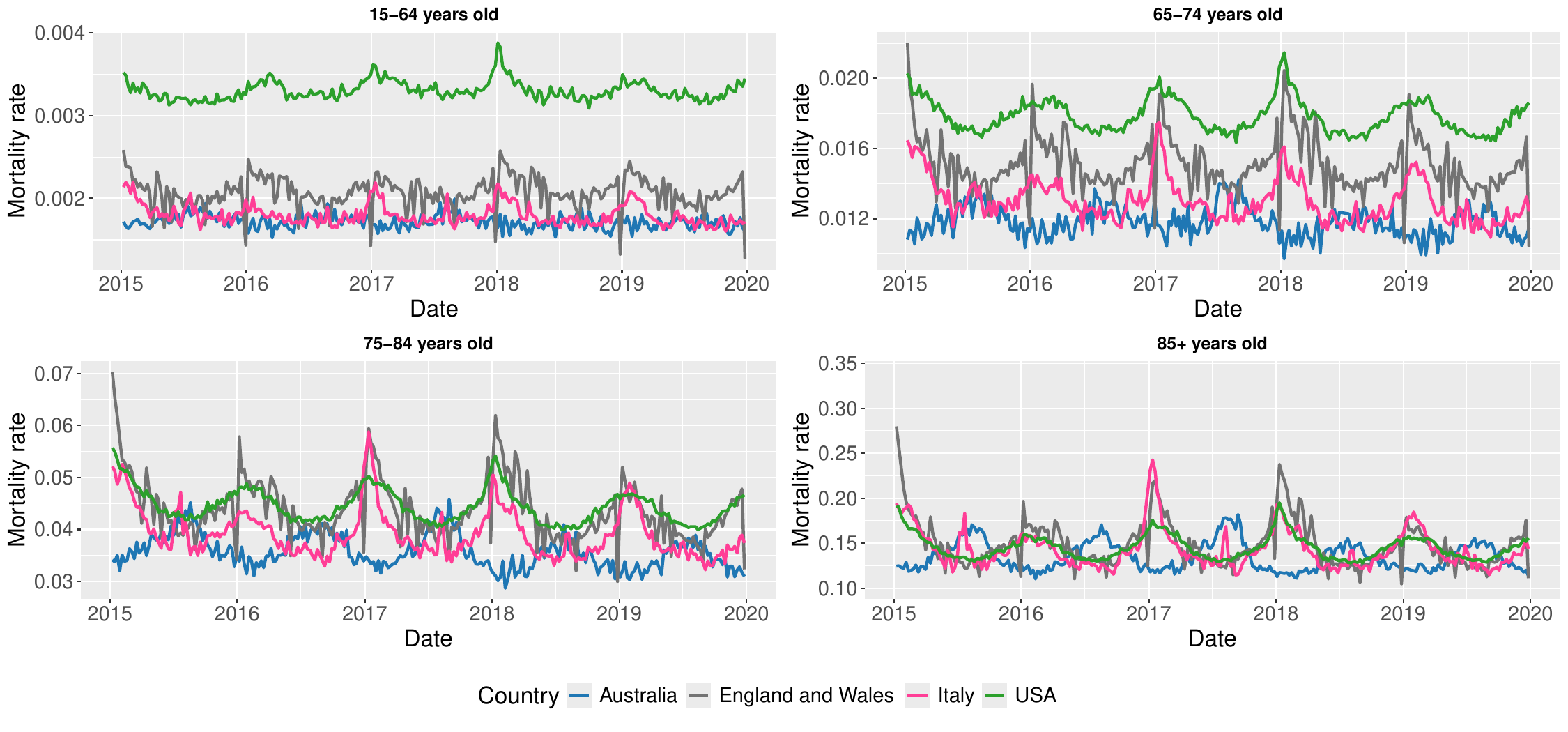}
\caption{2015--2019 mortality rates for the four representative countries by age groups}
\label{fig:4_country_data}
\end{figure}

The weekly mortality rates across four age groups for all 30 countries over the period 2015--2019 are presented in Appendix \ref{sec:Appendix_A}. However, the seasonal patterns are not easily discernible due to the high dimensions of the data. To facilitate a clearer examination of seasonal patterns, we present weekly mortality trends for four representative countries, Australia, England and Wales, Italy, and the US, in Figure \ref{fig:4_country_data}. The selection of these countries reflects both their geographic diversity, covering major regions such as Oceania, North America, and Europe, and the opportunity to compare contrasting seasonal patterns between the Northern and Southern Hemispheres. A recurring seasonal pattern is observed across all countries, with each cycle spanning approximately one year. Mortality rates tend to peak during the winter and decline in the summer, reflecting the influence of climate factors. As age increases, the seasonal pattern becomes smoother and more pronounced, suggesting a stronger and more apparent relationship between weather conditions and mortality among elderly individuals. This observation aligns with findings from prior research on weather-related mortality \citep{huang2015unusually}. Historical evidence indicates that events such as influenza outbreaks, heatwaves, and extreme cold spells have contributed to short-term fluctuations in mortality rates \citep{jdanov2021short}, and the impacts are more pronounced on individuals aged 55 and older \citep{li2022joint}.

We then take a closer look at the seasonal timing differences between the Southern and Northern Hemispheres. In Australia, peak mortality rates occur around the middle of the year, corresponding to the winter in the Southern Hemisphere. In contrast, Italy, England and Wales, as well as the US, exhibit peak mortality at the start of the year, reflecting their winter seasons in the Northern Hemisphere. Notably, beyond the typical U-shaped annual mortality pattern, a W-shaped seasonal trajectory appears in certain countries during specific years. This pattern reflects two distinct mortality peaks within a year: a primary and more pronounced peak in winter, commonly associated with cold waves, and a secondary, smaller summer peak likely driven by heatwaves. The W-shaped seasonal pattern is particularly evident among older age groups (75--84 and 85+ years old) and is especially pronounced in Italy and England and Wales.

\section{Methodology}
\label{sec:Method}
In this section, we propose a gradient boosting framework, which integrates the multi-population stochastic mortality model. Section \ref{sec:LL} presents a brief description of the Li and Lee model, which serves as the weak learner within the gradient boosting framework. In Section \ref{sec:GBLL}, we first introduce the gradient boosting framework, outlining its key components and applications, and then present the proposed Gradient Boosted Li and Lee (GBLL) model in detail.

Throughout the paper, the observed mortality rate for country $j$, age group $x$, and time $t$, denoted by $\tilde{m}^j_{x,t}$, is defined as $\tilde{m}^j_{x,t} = \tilde{D}^j_{x,t} / \tilde{E}^j_{x,t}$, where $\tilde{D}^j_{x,t}$ denotes the observed number of deaths and $\tilde{E}^j_{x,t}$ is the corresponding exposure for country $j$, age group $x$, and time $t$. Further details on the construction of these 
quantities can be found in \cite{stmf2021short}. We use $m^j_{x,t}$ to denote the modelled mortality rate and  $\hat{m}^j_{x,t}$ to denote the predictor of the mortality rates. A model parameter with a hat is used to denote an estimated value of the parameter (e.g., $\hat{\theta}$ denotes the predicted value of a generic model parameter $\theta$).
Bold symbols represent matrices, and lowercase letters refer to their elements (e.g., $x_{i,j}$ denotes the element in row $i$ and column $j$ of the matrix $\bm X$).

\subsection{Li and Lee (LL) model}
\label{sec:LL}
\cite{li2005coherent} propose the multi-population version of the Lee--Carter model to simultaneously estimate and forecast the mortality rates of a group of populations with similar economic conditions and close-knit connections. The Lee--Carter model proposed by \cite{lee1992modeling} serves as the foundation for the estimation of the Li and Lee model, and its definition and estimation algorithm are summarised in Appendix \ref{sec:Appendix_B}. \cite{li2005coherent} develop an augmented common factor model: For age group $x\in\{1,...,N\}$, time $t\in\{1,...,T\}$, and country $j\in\{1,...,J\}$, 
\begin{equation*}
\log(m^{j}_{x,t})=A^j_x + b^p_x \kappa^p_t + b^j_x \kappa^j_t + \varepsilon^j_{x,t},
\end{equation*}
where $m_{x,t}^j$ is the mortality rate for age group $x$ and country $j$ at time $t$, $A^j_x$ is the age-specific mean of log mortality rate over time for country $j$ and $\varepsilon^j_{x,t}$ is the corresponding error term for age group $x$, time $t$ and country $j$. The common trend is captured by $b^p_x$ and $\kappa^p_t$, obtained by applying the Lee--Carter method for the combined group including all the populations to avoid long-term divergence in mean mortality forecasts. In addition, $b^j_x$ and $\kappa^j_t$ represent the shorter-term excessive rate of mortality change in country $j$ over the common trend.

The original estimation method proposed by \cite{li2005coherent} first extracts the age-specific mean of log mortality rates $A_x^j$. Then the common trend $b^p_x \kappa^p_t$ across all sub-populations is computed using the Singular Value Decomposition (SVD) on the pooled, mean-adjusted log mortality rates $\log(m^{j}_{x,t})-A_x^j$. Population-specific residuals are then analysed via the SVD to obtain additional components $b^j_x \kappa^j_t$ to capture the deviations from the common trend. However, to ensure more coherent and simultaneous modelling across populations with other benchmark models, we follow the product-ratio functional method proposed by \cite{hyndman2013coherent} with the order chosen to be 1, to estimate the Li and Lee model. The detailed estimation is outlined as follows. First, let $\bm{Y}^j$ denote the $T \times N$ (time $\times$ age group) 
matrix of the observed log mortality rates for country $j$, whose $(t,x)$-th element 
is given by $y^j_{x,t} = \log(\tilde{m}^j_{x,t})$. We then re-express the observed mortality rate as $\tilde{m}^{j}_{x,t} = \tilde{p}_{x,t}\tilde{r}^{j}_{x,t}$ where $\tilde{p}_{x,t}=({\tilde{m}^{1}_{x,t}\times...\times \tilde{m}^{J}_{x,t}})^{1/{J}}$ is the product term, calculated as the geometric mean of the observed mortality rate across all countries, and $\tilde{r}^{j}_{x,t}=\tilde{m}^{j}_{x,t}/\tilde{p}_{x,t}$ represents the ratio term for country $j$. We use $\bm{Y}^j$ to fit separate Lee--Carter models to the product and ratio terms, to get the common factors $\bm{\hat{\kappa}}^p_t, \bm{\hat{a}}^p_x, \bm{\hat{b}}^p_x$ and country-specific factors $\bm{\hat{\kappa}}^{j}_t, \bm{\hat{a}}^{j}_x, \bm{\hat{b}}^{j}_x$, respectively; see Appendix \ref{sec:Appendix_B} for the model specification $a_x$, $b_x$, and $\kappa_t$ of the Lee--Carter model. We then compute the overall age-specific mean as $\bm{\hat{A}}^j_x=\bm{\hat{a}}^p_x + \bm{\hat{a}}^{j}_x$ and the final predicted log mortality rate is given by $$\hat{y}_{x,t}^j=\log(\hat{m}^{j}_{x,t})=\hat{A}^j_x+\hat{b}^p_x \hat{\kappa}^p_t+\hat{b}^{j}_x \hat{\kappa}^{j}_t,$$ for age group $x$, time $t$ and country $j$. The complete estimation procedure is summarised in Algorithm \ref{arg:LL}.

\begin{algorithm}[ht]
\setstretch{1.7}
\caption{Estimation of the LL Model (Product-Ratio Method with order 1)}
\KwInput{$\bm{Y^j}$ for $j\in\{1,...,J\}$ (all with size $T \times N$)}
\label{arg:LL}

\KwOutput{$\hat{y}_{x,t}^j$ for $x\in\{1,...,N\}$, $t\in\{1,...,T\}$ and $j\in\{1,...,J\}$\\

\hspace{1.7cm}$\bm{\hat{\kappa}}^p_t=(\hat{\kappa}^p_1,...,\hat{\kappa}^p_T), \bm{\hat{a}}^p_x=(\hat{a}^p_1,...,\hat{a}^p_N)$ \textbf{and} $\bm{\hat{b}}^p_x=(\hat{b}^p_1,...,\hat{b}^p_N)$\\

\hspace{1.7cm}$\bm{\hat{\kappa}}^j_t=(\hat{\kappa}^{j}_1,...,\hat{\kappa}^{j}_T), \bm{\hat{a}}^{j}_x=(\hat{a}^{j}_1,...,\hat{a}^{j}_N)$ \textbf{and} $\bm{\hat{b}}^{j}_x=(\hat{b}^{j}_1,...,\hat{b}^{j}_N)$}

\KwProcedure{LL($\bm{Y^j}$)}
\begin{algorithmic}[1]
\State $\tilde{p}_{x,t} \gets({\tilde{m}^{1}_{x,t}\times...\times \tilde{m}^{J}_{x,t}})^{\frac{1}{J}}, \forall x, t$

\State Denote $\bm{P}$ by $\log(\tilde{p}_{x,t})$ in row $t$ and column $x$

\State $\tilde{r}^{j}_{x,t} \gets \frac{\tilde{m}^{j}_{x,t}}{\tilde{p}_{x,t}} , \forall x, t$ and $j$

\State Denote $\bm{R}^j$ by $\log(\tilde{r}^{j}_{x,t})$ in row $t$ and column $x$

\State $\bm{\hat{\kappa}}^p_t, \bm{\hat{a}}^p_x, \bm{\hat{b}}^p_x \gets \text{LC}(\textbf{P})$, $\forall x, t$

\State $\bm{\hat{\kappa}}^j_t, \bm{\hat{a}}^j_x, \bm{\hat{b}}^{j}_x \gets \text{LC}(\bm{R}^{j})$, $\forall x, t$ and $j$

\State $\bm{\hat{A}}^j_x=\bm{\hat{a}}^p_x + \bm{\hat{a}}^{j}_x$, $\forall x, j$

\State $\hat{y}_{x,t}^j=\log(\hat{m}^{j}_{x,t}) \gets \hat{A}^j_x+\hat{b}^p_x \hat{\kappa}^p_t+\hat{b}^{j}_x \hat{\kappa}^{j}_t$, $\forall x, t$ and $j$
\end{algorithmic}
\end{algorithm}

\subsection{Gradient Boosted Li and Lee (GBLL) model}
\label{sec:GBLL}
Given concerns about the capacity of existing stochastic mortality models to accurately capture seasonality of mortality rates, we adopt a gradient boosting framework to iteratively model different layers of mortality trends. Gradient boosting is an additive modelling technique that iteratively improves predictive accuracy by correcting residual errors from previous fits \citep{friedman2001greedy}. Due to its flexibility and predictive power, the gradient boosting method can capture complex, nonlinear relationships and enhance both model fitting and forecast accuracy. Consequently, this data-driven approach has become increasingly popular in mortality rate modelling \citep[see e.g.,][]{blanes2021design, asgari2022comparison, qiu2022interpretable, neshat2026effective}.

There are two main components in the gradient boosting framework, namely the weak learner and the associated learning rate. Traditionally, a weak learner refers to a simple model with limited predictive power that performs slightly better than random guessing, such as a shallow decision tree. More recently, boosting has been used with other weak learners, including generalised linear models \citep{buhlmann2007boosting}, generalised additive models \citep{tutz2006generalized} and copulas \citep{hans2023boosting}. By sequentially aggregating each fitted weak learner to the final model with a learning rate that controls the contribution of each weak learner to the final model, it results in a stronger and more comprehensive model with enhanced forecast accuracy.

In this paper, we propose to employ the Li and Lee model as the weak learner to leverage domain expertise in mortality modelling, enhancing the interpretability of the overall model. Gradient boosting seeks to construct an ensemble of weak learners defined as $\sum_{g=1}^{l}\gamma_g\bm{\hat{y}}_{x,g}^{j}$, where $\hat{\bm{y}}_{x,g}^{j}$ denotes the $g$-th predicted log mortality rate from the Li and Lee model for country $j$ and age group $x$, $\gamma_g$ represents the associated coefficient, and $l$ is the number of iterations. This procedure involves fitting each weak learner to the residuals and solving a one-dimensional optimisation problem to determine its contribution.

To estimate the GBLL model, we first let $\bm{Y}^j = (\bm{y}_1^j,..., \bm{y}_N^j)$ denote the input mortality matrix with dimension $T \times N$ (time $\times$ age group) for country $j$. In the first iteration, the parameter set $\{\bm{\hat{\kappa}}_{t,1}^p, \bm{\hat{a}}_{x,1}^p, \bm{\hat{b}}_{x,1}^p, \bm{\hat{\kappa}}_{t,1}^j, \bm{\hat{a}}_{x,1}^j, \bm{\hat{b}}_{x,1}^j\}$ is obtained by fitting the Li and Lee model to $\bm{Y}^j$. Let $\bm{\hat{Y}}_1^j=(\bm{\hat{y}}_{1,1}^j,...,\bm{\hat{y}}_{N,1}^j)$ denote the fitted value of the first iteration. The learning rate $\gamma_1$ is then determined by minimising the following quadratic loss function: 
\begin{equation*}
L_1(\gamma) =\sum_{j=1}^{J} \left\| \bm{Y}^j - \gamma \bm{\hat{Y}}_1^j \right\|_F^2,
\end{equation*}
where $\left\| \cdot \right\|_F$ represents the Frobenius norm that measures the overall magnitude of a matrix, i.e., $L_1$ can be equivalently written as
\begin{equation*}
L_1(\gamma)=\sum_{j=1}^{J}\sum_{x=1}^{N}(\bm{y}_{x}^j-\gamma \bm{\hat{y}}_{x,1}^j)'(\bm{y}_{x}^j-\gamma \bm{\hat{y}}_{x,1}^j).
\end{equation*}
In the $(g+1)$-th iteration for $g\ge 1$, the parameter set $\{\bm{\hat{\kappa}}_{t,{g+1}}^p, \bm{\hat{a}}_{x,{g+1}}^p, \bm{\hat{b}}_{x,{g+1}}^p, \bm{\hat{\kappa}}_{t,{g+1}}^j, \bm{\hat{a}}_{x,{g+1}}^j, \bm{\hat{b}}_{x,{g+1}}^j\}$ and the fitted value $\bm{\hat{Y}}^j_{g+1} = (\bm{\hat{y}}_{1,g+1}^j, \ldots, \bm{\hat{y}}_{N,g+1}^j)$
are obtained by fitting the Li and Lee model to residual $\bm{E}^j_g = (\bm{e}_{1,g}^j, ..., \bm{e}_{N,g}^j)$ from the $g$-th iteration, defined by: 
\begin{equation*}
\bm{E}^j_{g} =
\begin{cases}
\bm{Y}^j - \gamma_1 \bm{\hat{Y}}^j_{1} = (\bm{e}_{1,1}^j,..., \bm{e}_{N,1}^j)=(\bm{y}_{1}^j - \gamma_1 \bm{\hat{y}}_{1,1}^j, ..., \bm{y}_{N}^j - \gamma_1 \bm{\hat{y}}_{N,1}^j), & \text{if } g = 1, \\
\bm{E}^j_{g-1} - \gamma_g \bm{\hat{Y}}^j_{g} = (\bm{e}_{1,g}^j,..., \bm{e}_{N,g}^j)=(\bm{e}_{1,g-1}^j - \gamma_g \bm{\hat{y}}_{1,g}^j, ..., \bm{e}_{N,g-1}^j - \gamma_g \bm{\hat{y}}_{N,g}^j), & \text{if } g \geq 2.
\end{cases}
\end{equation*}
The learning rate $\gamma_{g}$ for $g\ge 2$ is determined by minimising the following quadratic loss function:
\begin{equation*}
L_g(\gamma) =\sum_{j=1}^{J} \left\| \bm{E}_{g-1}^j - \gamma \bm{\hat{Y}}_g^j \right\|_F^2.
\end{equation*}

The algorithm terminates when (a) all 120 residual series (30 countries, 4 age groups each) have Ljung--Box test p--values of at least 0.05 \citep{ljung1978measure}, or (b) the maximum in-sample fitting MAPE across all age groups, countries and training periods falls below a small threshold. For condition (a), the p--value $p_{x,g}^j$ for each age group $x$, country $j$ and iteration $g$ is given by $1 - F_{\chi^2_{52}}\left(T(T+2) \sum_{k=1}^{52} (\rho_{k,x,g}^{(j)})^2/(T-k)\right)$, where $F_{\chi^2_{52}}(\cdot)$ represents the Cumulative Distribution Function (CDF) of a chi-squared distribution with 52 degrees of freedom, $T$ is the whole training period, lag is chosen at 52, and $\rho_{k,x,g}^{(j)}$ is the sample autocorrelation at lag $k$ for the residual $\bm{e}_{x,g}^j$ for country $j$, age group $x$ and the $g$-th iteration. For condition (b), the maximum in-sample fitting MAPE across all age groups $x\in\{1,...,N\}$, countries $j\in\{1,...,J\}$, and training periods $t\in\{1,...,T\}$ is defined as (see Section \ref{sec:forecasting} for more details on the definition of MAPE):
\begin{equation*}
    \underset{x,j,t}{\max}\left(\mathrm{MAPE}_{x,j,t}\right)=\underset{x,j,t}{\max}\left(\frac{|\hat{m}^j_{x,t}-\tilde{m}^j_{x,t}|}{\tilde{m}^j_{x,t}}\right).
\end{equation*}
The algorithm stops if the maximum in-sample fitting MAPE is less than $0.001$. 
This constraint prevents unnecessary iterations once a sufficiently accurate fit is reached and avoids reducing the fitting error to values that may reflect overfitting rather than meaningful improvement. The 0.001 threshold aligns with common optimisation tolerances, generally ranging from $10^{-6}$ to $10^{-3}$ \citep{nocedal2006numerical}. It suggests that the residuals contain no remaining systematic patterns and the model adequately captures the underlying data structure. We set the maximum number of boosting iterations to $G = 50$. A detailed summary of this procedure is provided in Algorithm \ref{arg:GBLL}. The fitted log mortality rates are obtained as follows: For age group $x\in\{1,...,N\}$, time $t\in\{1,...,T\}$, and country $j\in\{1,...,J\}$,
\begin{equation*}
\log(\hat{m}^{j}_{x,t}) = \sum_{g=1}^{l} \gamma_g\left(\hat{a}^p_{x,g}+\hat{b}^p_{x,g}\hat{\kappa}^p_{t,g}+\hat{a}^j_{x,g}+\hat{b}^j_{x,g}\hat{\kappa}^j_{t,g}\right). 
\end{equation*}

\begin{algorithm}[!ht]
\setstretch{1.5}
\caption{Estimation of the GBLL Model}
\label{arg:GBLL}
\KwInput{$\bm{Y}^j$ for $j\in\{1,...,J\}$ (all with size $T \times N$)}

\KwOutput {$\bm{\hat{\kappa}}_{t,l}^p, \bm{\hat{a}}_{x,l}^p, \bm{\hat{b}}_{x,l}^p$ and $\gamma_l$ for $x\in\{1,...,N\}$, $t\in\{1,...,T\}$ and $l\in\{1,...,G\}$\\

\hspace{1.7cm}$\bm{\hat{\kappa}}_{t,l}^j, \bm{\hat{a}}_{x,l}^j, \bm{\hat{b}}_{x,l}^j$ for $x\in\{1,...,N\}$, $t\in\{1,...,T\}$,  $l\in\{1,...,G\}$ and $j\in\{1,...,J\}$}

\KwProcedure{GBLL($\bm{Y}^j$)}
\begin{algorithmic}[1]
\State $\bm{\hat{y}}_{x,1}^j, \bm{\hat{\kappa}}_{t,1}^p, \bm{\hat{a}}_{x,1}^p, \bm{\hat{b}}_{x,1}^p, \bm{\hat{\kappa}}_{t,1}^j, \bm{\hat{a}}_{x,1}^j, \bm{\hat{b}}_{x,1}^j \gets \text{LL}(\bm{Y}^j)$, $\forall x, t$ and $j$

\State $L_1 \gets \sum_{j=1}^{J}\sum_{x=1}^{N}(\bm{y}_{x}^j-\gamma \bm{\hat{y}}_{x,1}^j)'(\bm{y}_{x}^j-\gamma \bm{\hat{y}}_{x,1}^j)$

\State $\gamma_1 =  \underset{\gamma}{\arg\min} L_1$

\State $\bm{e}_{x,1}^j \gets\bm{y}_{x}^j - \gamma_1 \bm{\hat{y}}_{x,1}^j$, $\forall x, j$

\State $\bm{E}^j_1 \gets (\bm{e}_{1,1}^j, ..., \bm{e}_{N,1}^j)$,  $\forall j$

\State $g \gets 1$

\State
\While{$g\leq G$}{
    $p_{x,g}^j \gets$ $1 - F_{\chi^2_{52}}\left(T(T+2) \sum_{k=1}^{52} \frac{(\rho_{k,x,g}^{(j)})^2}{T-k}\right)$, the p--value of $\bm{e}_{x,g}^j$ under the Ljung--Box test for country $j$ and age group $x$\;
    \If{$g = G \;\mathbf{or}\; \underset{x,j}{\min} \; p_{x,g}^j \ge 0.05 \;\mathbf{or}\; \underset{x,j,t}{\max} \; \frac{|\hat{m}^j_{x,t}-\tilde{m}^j_{x,t}|}{\tilde{m}^j_{x,t}} < 0.001$}{
    $l \gets g$\;
    \textbf{break}\;
    }
    \Else{
    $\bm{\hat{y}}_{x,g+1}^j$, $\bm{\hat{\kappa}}_{t,g+1}^p$, $\bm{\hat{a}}_{x,g+1}^p$, $\bm{\hat{b}}_{x,g+1}^p$, $\bm{\hat{\kappa}}_{t,g+1}^j$, $\bm{\hat{a}}_{x,g+1}^j$, $\bm{\hat{b}}_{x,g+1}^j \gets \text{LL}(\bm{E}^j_g)$, $\forall x,t$ and $j$\;

    $L_{g+1} \gets \sum_{j=1}^{J}\sum_{x=1}^{N}(\bm{e}_{x,g}^j-\gamma \bm{\hat{y}}_{x,g+1}^j)'(\bm{e}_{x,g}^j-\gamma \bm{\hat{y}}_{x,g+1}^j)$\;

    $\gamma_ {g+1}=  \underset{\gamma}{\arg\min} L_{g+1}$\;

    $\bm{e}_{x,g+1}^j \gets\bm{e}_{x,g}^j - \gamma_{g+1} \bm{\hat{y}}_{x,g+1}^j$, $\forall x,j$\;

    $\bm{E}^j_{g+1} \gets (\bm{e}_{1,g+1}^j, ..., \bm{e}_{N,g+1}^j)$,  $\forall j$\;

    $g \gets g+1$
    }
}

\end{algorithmic}
\end{algorithm}

To ensure consistency in seasonal mortality patterns across countries, particularly between the hemispheres, we address the phase reversal in seasonal cycles observed in weekly mortality data. Specifically, winter-related mortality peaks occur in opposite calendar months across hemispheres. During model estimation, we therefore apply a temporary monotonic reciprocal transformation (i.e., $\tilde{m}_{x,t}^{j*}=1/\tilde{m}_{x,t}^j$ where $\tilde{m}_{x,t}^{j*}$ and $\tilde{m}_{x,t}^j$ are the transformed and raw observed mortality rates, respectively) to the Southern Hemisphere mortality rates (Australia and New Zealand) to align the phase of their seasonal patterns with those of Northern Hemisphere countries. This transformation is used solely for estimation purposes to facilitate the identification of a common seasonal component within the multi-population framework. After forecasting, all series are transformed back to their original scale (i.e., $\hat{m}_{x,t}^{j}=1/\hat{m}_{x,t}^{j*}$ where $\hat{m}_{x,t}^{j}$ is the fitted value of the mortality rate and $\hat{m}_{x,t}^{j*}$ is the fitted value of the transformed mortality rate), ensuring that the resulting mortality forecasts remain fully interpretable.

Forecasting mortality rates primarily relies on projecting both the common and country-specific time trend component $\kappa_t$. In classical mortality modelling, $\kappa_t$ is often assumed to follow a random walk with drift. For high-frequency weekly mortality data, a natural extension is to employ a seasonal-ARIMA (SARIMA) model to account for annual seasonality. However, SARIMA models require seasonal differencing at lag 52, which effectively removes a full year of observations from the already short five-year sample (260 weeks), potentially leading to unstable parameter estimates and unreliable forecasts. Moreover, SARIMA with a long seasonal period is computationally intensive, especially in a multi-population gradient boosting framework. To overcome this limitation, we adopt a hybrid forecasting approach in which seasonal patterns are modelled deterministically using Fourier regressors \citep{serfling1963methods}, while the residual stochastic dynamics are captured by a non-seasonal ARIMA$(p,d,q)$ process. This strategy allows us to retain the full sample for estimation while flexibly capturing both annual and biannual seasonal cycles. Specifically, $\kappa_t$ is modelled as
\begin{equation*}
\kappa_t = \beta_1 \sin\!\big(2\pi w(t)\big) + \beta_2 \cos\!\big(2\pi w(t)\big) + \beta_3 \sin\!\big(4\pi w(t)\big) + \beta_4 \cos\!\big(4\pi w(t)\big) + \text{ARIMA}(p,d,q),
\end{equation*}
where $w(t)\in[0,1)$ denotes the fractional part of the calendar year corresponding to time $t$. There are four Fourier regressors, including both sine and cosine functions to capture both the annual and biannual cycles. The statistical significance of the Fourier components is assessed using a Wald test on the corresponding regression coefficients \citep{wald1943tests}. Specifically, for each Fourier term, the null hypothesis that its coefficient is equal to zero is tested using the associated z-statistic $z_i = \hat{\beta}_i / \mathrm{SE}(\hat{\beta}_i)$, for $i = 1,2,3,4$, along with the corresponding p--value obtained from the \texttt{coeftest} function in \texttt{R}. Under the large-sample approximation, the test statistic follows $Z \sim \mathcal{N}(0,1)$. A Fourier regressor is deemed statistically significant at the 5\% level if its p--value is less than 0.05. If none of the Fourier components satisfy this criterion, the forecasting procedure relies solely on the ARIMA$(p,d,q)$ model. Therefore, standing on the final observed time $T$, 
the final $h$-step-ahead forecasts of the log mortality rate is given by: For age group $x\in\{1,...,N\}$, $j\in\{1,...,J\}$,
\begin{equation*}
\log(\hat{m}^{j}_{x,T+h}) = \sum_{g=1}^{l} \gamma_g\left(\hat{a}^p_{x,g}+\hat{b}^p_{x,g}\hat{\kappa}^p_{T+h|T,g}+\hat{a}^j_{x,g}+\hat{b}^j_{x,g}\hat{\kappa}^j_{T+h|T,g}\right). 
\end{equation*}

There are alternative choices for each component of the GBLL model, including the loss function, stopping criterion, models used to fit and forecast the time trend $\kappa_t$, and seasonal adjustments for countries in the Southern Hemisphere. Detailed descriptions of these alternatives are provided in Section A of the Supplementary Material. Nevertheless, our main conclusions remain unchanged, and the proposed GBLL model is shown to be insensitive to the choice of individual components.

\section{Empirical results}
\label{sec:results}
This section evaluates both the in-sample fitting (Section \ref{sec:fitting}) and out-of-sample forecast performance (Section \ref{sec:forecasting}) of the proposed gradient boosted Li and Lee model, in comparison with the baseline Li and Lee model (see Section \ref{sec:LL}), the Hyndman--Booth--Yasmeen (HBY) model, the Vector Auto-Regressive (VAR) model, and the multi-population Global Vector Auto-Regressive (GVAR) model. An overview of the HBY model is given in Appendix \ref{sec:HBY} and Appendix \ref{sec:VAR} contains a detailed analysis based on the VAR and GVAR model.

To mitigate the impact of stochastic variability in the model fitting and forecast process, we adopt a 10-fold expanding window approach. In this setup, the starting point of the estimation window is fixed at the initial point of the dataset, the second week of January 2015. The endpoint is incrementally advanced from the final week of March 2018 to the final week of December 2018, increasing by one month at each step. This results in a progressively larger training window over successive iterations. The forecast horizon is held constant at 52 weeks, and forecast accuracy is computed monthly across all age groups and all countries. By reducing the estimation bias through repeated sampling, this approach yields more reliable and stable results.

\subsection{Fitting performance}
\label{sec:fitting}
Empirically, the GBLL algorithm terminates due to condition (b), typically after around 8 iterations. The maximum in-sample fitting error across all horizons, countries, and age groups is below 0.001, indicating that the model has captured all underlying significant patterns and meaningful variations in the data. This adequacy can be further assessed by examining the number of residual series exhibiting white-noise behaviour across age groups and countries using the Ljung--Box test at a 5\% significance level \citep{ljung1978measure}.

For each expanding window, we count the number of residual series identified as white noise (out of 120, corresponding to 30 countries and 4 age groups). The average number of white-noise residuals across the 10 expanding windows show that the LL and HBY models produce 54 and 15 white-noise residuals, respectively, while the proposed GBLL model (with 8 iterations) produces 73. This indicates that the GBLL model captures nearly all significant and meaningful patterns in the mortality data, as also reflected by the small maximum in-sample fitting error (below 0.001). In contrast, the HBY model fails to capture some structures, particularly for the 65--74 and 75--84 age groups, while the LL model performs better but still leaves more residual structure unexplained. The higher number of white-noise residuals in the GBLL model, together with the relatively small fitting error, demonstrates its superior ability to extract the underlying trends and seasonality across countries and age groups. It is worth noting that our comparison focuses on the HBY model with order 6; alternative specifications may perform differently, but exploring the full range of configurations is beyond the scope of this paper.

\begin{figure}[ht]
\centering
\includegraphics[width=1\textwidth]{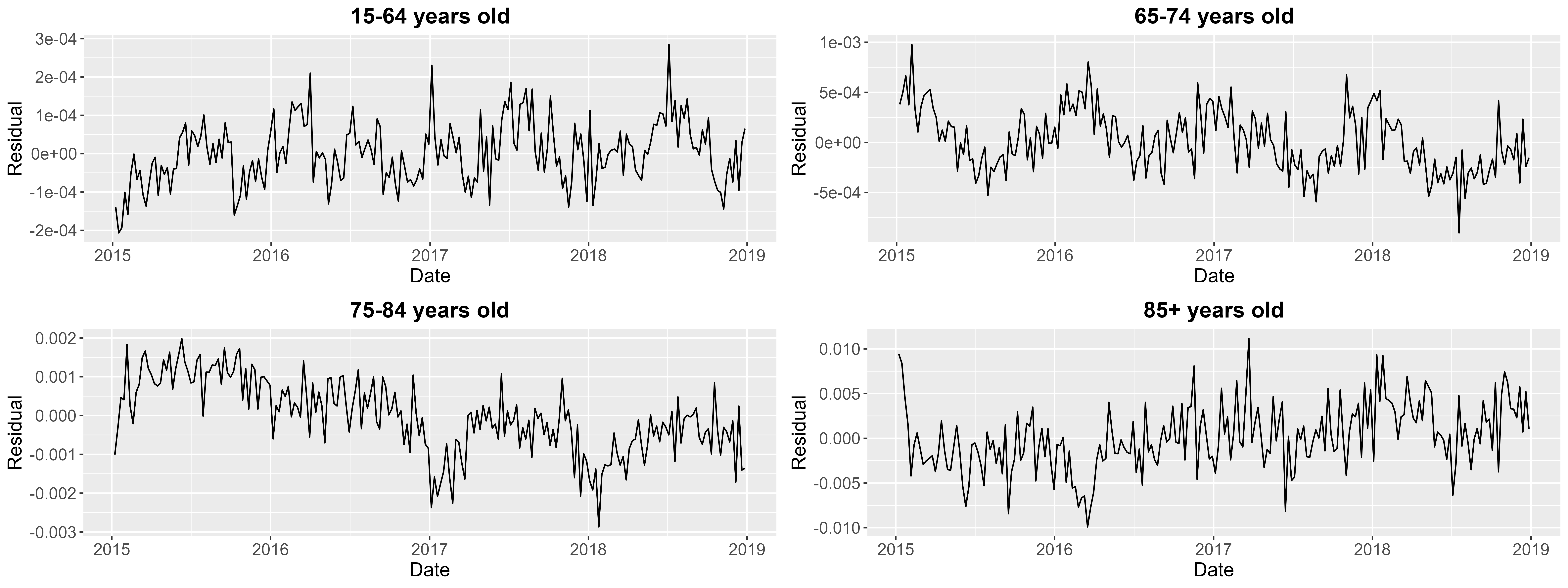}
\caption{The residual plot of Canada under the LL model}
\label{fig:canada_residual_LL}
\end{figure}

\begin{figure}[htbp]
\centering
\includegraphics[width=1\textwidth]{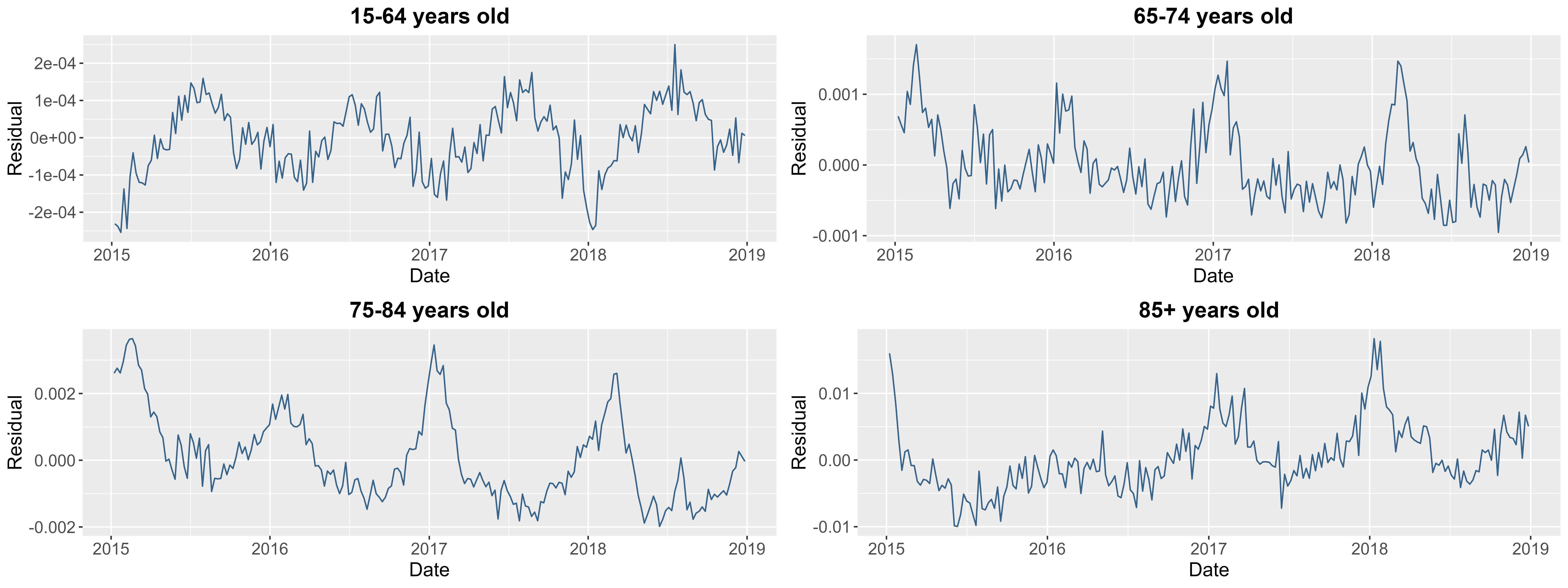}
\caption{The residual plot of Canada under the HBY model}
\label{fig:Canada_residual_HBY}
\end{figure}

Looking closely at those residuals that are not white noise, there remain either seasonal cyclical patterns or clear downwards/upwards sloping trends or both. This means that, given the complexity and high dimensions of the dataset, it is not suitable to use the existing methods, as they fail to capture all the patterns. For instance, in the case of Canada, residual plots from both the LL and HBY models reveal remaining seasonal cycles and trends. Figure \ref{fig:canada_residual_LL} shows clear seasonal cycles for the 65--74 years old and a downwards sloping trend for the 75--84 years old, under the LL model. Similarly, Figure \ref{fig:Canada_residual_HBY} gives the residual plot of Canada under the HBY model and there are clear seasonal cyclical patterns for all the four age groups and a slightly decreasing trend for the 65--74 years old. Notably, the 15--64 age group exhibits an inverse seasonal pattern. This may be due to the model imposing seasonal structure to accommodate for other age groups, even though the raw data shows little seasonal variation. Additionally, the residuals for individuals aged 85 and over are approximately around $\pm 0.01$, clearly indicating that both the LL and HBY models struggle to capture the underlying patterns for this age group. In contrast, the residuals across all four age groups under the GBLL model exhibit white-noise behaviours, indicating improved model adequacy. Additional visualisations of in-sample fits for the GBLL model and the benchmark models for selected countries are provided in Section C.1 of the Supplementary Material.

\vspace{0.1in}

\begin{figure}[htbp]
\centering
\includegraphics[width=1\textwidth]{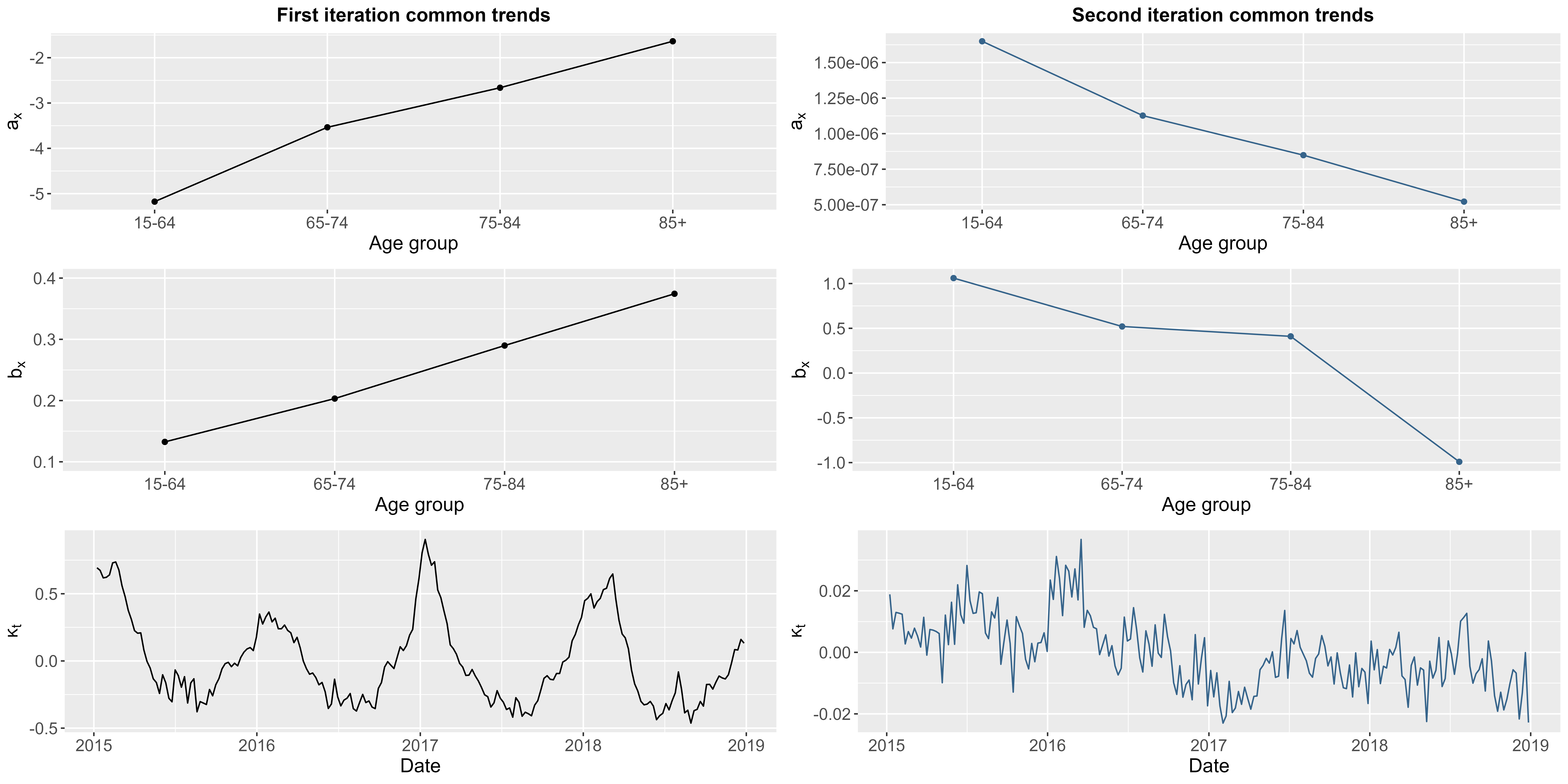}
\caption{Components of the common trends under the GBLL model: first iteration (left) and second iteration (right)}
\label{fig:GBLL_components_common}
\end{figure}

Finally, Figure \ref{fig:GBLL_components_common} illustrates the contribution of each component at different stages of the model fitting process. We present the decomposition of the GBLL model for the first two iterations here, focusing on the common trends. The country-specific components breakdown are provided in Appendix \ref{sec:Appendix_C}. The first column (left) displays, from top to bottom, the estimated age-specific intercepts ($\hat{a}_x$), age loadings ($\hat{b}_x$), and the time-varying index ($\hat{\kappa}_t$) from the first iteration. The second column (right) presents the corresponding components from the second iteration.

\begin{itemize}
\item For the age-specific intercepts $\hat{a}_x$, the first iteration captures a generally increasing trend. It highlights the vulnerability of older people, resulting in higher average mortality rates. In the second iteration, a predominantly decreasing pattern is observed. Importantly, the magnitude of these adjustments is on the order of $10^{-6}$, suggesting that the primary age-related mortality pattern is effectively captured in the first iteration.

\item For the age loadings $\hat{b}_x$, the first iteration exhibits an upward trend, indicating the greater exposure experienced by elderly individuals to higher mortality rates. In the second iteration, a declining trend is observed. This may be related to the original smoother and more pronounced seasonal patterns typically exhibited by older age groups. As these dominant effects are captured in the first iteration, only a small portion of the residual variation remains to be explained. Additionally, some spikes exist in the 15--64 and 85+ age groups, indicating that this iteration primarily captures residual effects concentrated in the extreme age cohorts. 

\item For the common time trend component $\hat{\kappa}_t$, the first iteration effectively captures the dominant seasonal cycles. The second iteration provides a refinement by identifying smaller-scale cyclical patterns with short-term fluctuations and a mild downward trend, indicative of gradual improvements in mortality rates over time. A prominent W-shaped seasonal pattern is evident in the first iteration across all years, arising from the aggregation of mortality dynamics across all countries. This aggregation enhances structural features that may be weakly present but commonly shared across populations, thereby making the W-shape (with dual seasonal peaks) more pronounced. Additionally, an M-shaped pattern is discernible, particularly around early 2015 and 2018. This may be attributed to the reciprocal transformation applied to the Southern Hemisphere countries, where a secondary mortality peak, likely associated with heatwaves, typically occurs at the beginning of the calendar year.
\end{itemize}

\subsection{Forecast performance with expanding windows}
\label{sec:forecasting}
To evaluate the forecast performance, we adopt the Mean Absolute Percentage Error (MAPE) as the error measure to quantify the average magnitude of the percentage deviation between forecasted and actual values. MAPE is scale-independent, allowing direct comparison of forecast accuracy across countries with different mortality levels. It has been widely used in the mortality forecasting literature as a standard metric for evaluating predictive performance \citep[see e.g.,][]{mcnown1992forecasting, neves2017five, li2019forecast, tsai2021incorporating, qiao2024machine}. We first define the set $H= \{4, 9, 13, 17, 22, 26, 30, 35, 39, 43, 48, 52\}$, which corresponds to the cumulative number of weeks until the end of each month, thereby facilitating the conversion from weekly to monthly forecast horizons. For month $h\in\{1,...,12\}$, expanding window $r\in\{0,...,9\}$, the MAPE for $h$-step-ahead forecasts across all age groups, countries and expanding windows is given by:
\begin{equation*}
\mathrm{MAPE}_h=\frac{1}{4\times H_{(h)} \times 30 \times 10}\sum_{x=1}^4\sum_{u=1}^{H_{(h)}}\sum_{j=1}^{30}\sum_{r=0}^{9}\frac{|\hat{m}^j_{x,169+u+H_{(r)}}-\tilde{m}^j_{x,169+u+H_{(r)}}|}{\tilde{m}^j_{x,169+u+H_{(r)}}},
\end{equation*}
where $H_{(i)}$ represents the $i$-th order statistic of the set $H$ with $H_{(0)}=0$, and 169 is the number of weeks in the first training sample. An additional scale-invariant error metric, the Mean Absolute Scaled Error (MASE), is also employed, and leads to comparable conclusions to those obtained using MAPE. Full results and discussions are provided in Section B of the Supplementary Material.

In addition to the factor-based LL and HBY models, we include further comparisons of forecast performance using Vector Auto-Regressive (VAR) models. The specifications of both the single-population VAR and the multi-population GVAR models are provided in Appendix \ref{sec:VAR}. Table \ref{tab:forecast_MAPE_whole} presents the mean MAPE of out-of-sample forecasts over the forecast horizon $h$ from 1 to 12 months (scaled by a factor of 100). For each horizon, the lowest mean MAPE, indicating the best-performing model, is highlighted in bold. The results indicate that, based on the STMF dataset, the proposed GBLL model outperforms the two benchmark factor models and the two Vector Auto-Regressive models across all forecast horizons. This superior performance is achieved by the iterative structure of the gradient boosting framework, which enables the model to sequentially identify and incorporate multiple layers of complex patterns. The comparatively weaker performance of VAR and GVAR models can be attributed to their generic time series nature and the differencing required to ensure stationarity. In contrast to factor models (LL, HBY and GBLL), which are specifically designed to capture and forecast long-run mortality trends, VAR-type models are typically estimated on differenced series and are therefore better suited to analysing short-run dynamics, such as spillover effects and shock transmission as characterised by impulse response functions. Moreover, the relatively short time span of mortality data exacerbates the curse of dimensionality in VAR and GVAR models, leading to increased estimation uncertainty. Accordingly, subsequent detailed comparisons of model performance will focus only on the class of factor-based models.

\begin{table}[htbp]
  \centering
  \caption{Mean MAPE of out-of-sample forecasts across 30 countries ($\times 100$ scale)}
    \begin{tabular}{cccccc}
    \toprule
    \makebox[1.5cm]{\textbf{$h$}} & \makebox[2cm]{\textbf{LL}} & \makebox[2.3cm]{\textbf{HBY}} & \makebox[2.3cm]{\textbf{GBLL}} & \makebox[2.3cm]{\textbf{VAR}} & \makebox[2.3cm]{\textbf{GVAR}}\\
    \midrule
    1     & 5.436 & 5.985 & \textbf{5.278} & 8.107 & 10.637 \\
    2     & 5.779 & 6.512 & \textbf{5.620} & 8.707 & 11.063 \\
    3     & 5.746 & 6.537 & \textbf{5.582} & 8.991 & 11.540 \\
    4     & 5.774 & 6.547 & \textbf{5.602} & 9.259 & 12.017 \\
    5     & 5.846 & 6.553 & \textbf{5.671} & 9.578 & 12.620 \\
    6     & 5.889 & 6.563 & \textbf{5.718} & 9.783 & 13.057 \\
    7     & 5.923 & 6.610 & \textbf{5.753} & 9.958 & 13.446 \\
    8     & 5.998 & 6.698 & \textbf{5.829} & 10.134 & 13.847 \\
    9     & 6.048 & 6.752 & \textbf{5.878} & 10.239 & 14.095 \\
    10    & 6.085 & 6.758 & \textbf{5.914} & 10.290 & 14.259 \\
    11    & 6.096 & 6.745 & \textbf{5.922} & 10.288 & 14.306 \\
    12    & 6.091 & 6.726 & \textbf{5.911} & 10.265 & 14.305 \\
    \bottomrule
    \end{tabular}%
  \label{tab:forecast_MAPE_whole}%
\end{table}%

\begin{figure}[ht]
\centering
\includegraphics[width=1\textwidth, height=0.28\textheight]{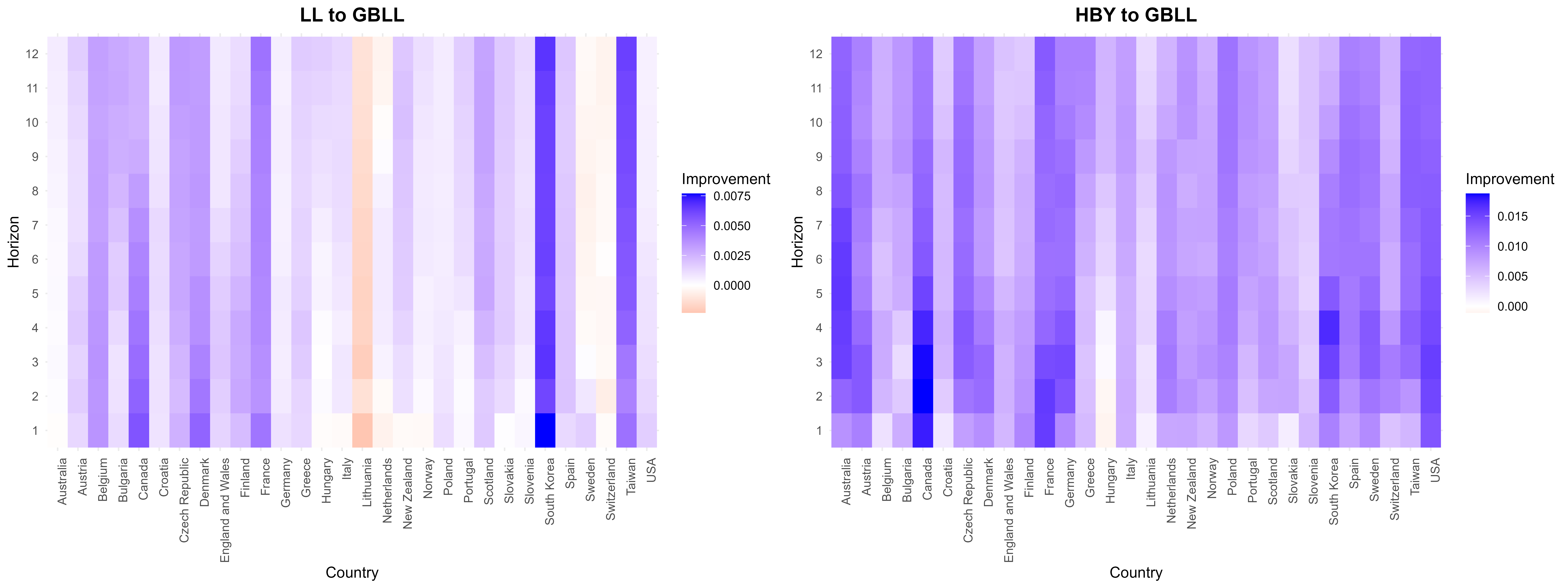}
\caption{Improvement in MAPE from LL to GBLL (left) and from HBY to GBLL (right)}
\label{fig:Improvement_to_GBLL}
\end{figure}

To gain a better understanding of the enhanced forecast accuracy achieved by the proposed GBLL model, Figure \ref{fig:Improvement_to_GBLL} presents the improvement in MAPE across countries and forecast horizons. Compared to the standard LL model, the GBLL model enhances forecast accuracy. Substantial gains are observed in countries such as South Korea and Taiwan, while Lithuania is slightly better under the LL model. Relative to the HBY model, GBLL again shows superior performance in nearly all cases, except for Hungary in the first two forecast months. Overall, the GBLL model achieves the highest accuracy across the three models evaluated. Additional visualisations of out-of-sample forecasts for the GBLL model and the benchmark models for selected countries are provided in Section C.2 of the Supplementary Material. Interestingly, the LL model outperforms the HBY model despite its simpler structure. A likely explanation is that the use of only four age groups, as opposed to single-year-of-age data for which the HBY model is designed, reduces the number of principal components needed to explain the variation in data. Consequently, the additional principal components may contribute little beyond the first. Therefore, given the complexity and heterogeneity of mortality data across 30 countries, simpler or progressive approaches such as the gradient boosting framework, may be better suited to capturing shared and country-specific patterns than more complex one-step models. We have conducted an additional analysis to evaluate forecast performance across age groups, with a detailed discussion provided in Section D of the Supplementary Material. Overall, the proposed GBLL model demonstrates superior forecast accuracy for most age groups, particularly among younger cohorts, while the simpler LL model performs comparably for the oldest age group, likely due to its smoother mortality patterns.

In addition to evaluating point forecast accuracy, we assess each model’s ability to quantify predictive uncertainty via marginal calibration of 95\% prediction intervals \citep{czado2009predictive}. For each model, 100 forecast paths for $\kappa_
t$ are simulated, and prediction intervals for $\hat{m}^j_{x,t}$ are constructed from the empirical 2.5th and 97.5th percentiles, denoted $\hat{p}^j_{2.5,x,t}$ and $\hat{p}^j_{97.5,x,t}$. Marginal calibration evaluates whether the nominal probability of the interval aligns with the empirical coverage, i.e., the proportion of observed outcomes that fall within the interval. Formally, for week $h_w \in \{1,...,52\}$, expanding window $r \in \{0,...,9\}$, age group $x \in \{1,...,4\}$, and country $j \in \{1,...,30\}$, the marginal calibration $f_{[\hat{p}^j_{2.5,x,t},\hat{p}^j_{97.5,x,t}]}$ is given by:
{\footnotesize
\begin{equation*}
f_{[\hat{p}^j_{2.5,x,t},\hat{p}^j_{97.5,x,t}]} = \frac{1}{52 \times 10 \times 4 \times 30} \sum_{h_w=1}^{52} \sum_{r=0}^{9} \sum_{x=1}^{4} \sum_{j=1}^{30} \mathds{1}\Big(\hat{p}^j_{2.5,x,169+h_w+H_{(r)}} \le \tilde{m}^j_{x,169+h_w+H_{(r)}} \le \hat{p}^j_{97.5,x,169+h_w+H_{(r)}} \Big),
\end{equation*}}where $\mathds 1$ denotes the indicator function. Mean empirical marginal calibration rates show that the LL, HBY, and GBLL models achieve 90\%, 94\%, and 96\% coverage, respectively. These results indicate that the GBLL model is the most properly calibrated, producing prediction intervals that reliably capture realised mortality rates and providing improved uncertainty quantification alongside superior point forecast performance.

\section{An experiment on clustering}
\label{sec:clustering}
Clustering is an unsupervised learning technique that partitions data into distinct groups based on similarity, aiming to ensure that observations within the same cluster are more alike than those in different clusters \citep{rokach2005clustering}. Clustering can offer an alternative way to group data based on characteristics such as geographic, socio-economic, or seasonal factors, and can therefore be used to assess the sensitivity of the model to the underlying data structure. Evaluating model performance across these clusters provides insight into the extent to which the model responds to structural heterogeneity. Limited variation in performance indicates stability, whereas substantial discrepancies may signal sensitivity to differences across groups. Ensuring homogeneity within clusters enhances coherence in sub-population forecasts, though achieving this is more challenging in cross-country comparisons due to diverse environmental, social and political backgrounds.

Clustering techniques are increasingly employed in mortality rate modelling to enhance the interpretability and accuracy of forecasts. \cite{hatzopoulos2013common} aim to construct common mortality trends across multiple populations with similar mortality experiences, enabling more coherent forecasts. They apply fuzzy c-means clustering to the main time effects, which serve as a summary of mortality dynamics. Similarly, \cite{leger2021can} adopt a functional data analysis framework to mix countries and ages in the modelling process. \cite{tsai2021incorporating} investigate age-specific mortality trends and propose that grouping ages with similar temporal patterns can improve forecast performance.

\subsection{Set up}
\label{sec:3_clustering_methods}
Clustering typically involves four key components: the data input, clustering algorithm, distance measure, and the optimal number of clusters \citep{yin2024rapid}. In this paper, we first extract the time trend component $\kappa_t$ for each country using the Lee--Carter model. We then consider three different variations of the raw time series as the data input to investigate population similarity, while keeping the remaining three clustering components fixed.

Clustering algorithms can be broadly classified into several categories, including partitional, hierarchical, density-based, grid-based, and model-based approaches \citep{ghosal2020short}. In this study, we adopt the K-means algorithm, a representative method within the partitional clustering framework, which minimises within-cluster sum of squares through an iterative process of centroid updating and point re-assignment \citep{lloyd1982least}. The accompanying method to find the optimal number of clusters is the elbow method, which plots the inertia (within-cluster sum of squares) against a range of values for the number of clusters \citep{thorndike1953belongs}. Inertia is calculated as:
\begin{equation*}
\text{Inertia} = \sum_{q=1}^{Q} \sum_{\mathbf{x} \in C_q}  (\mathbf{x}-\bm{\mu}_q)'(\mathbf{x}-\bm{\mu}_q), 
\end{equation*}
where $Q$ is the total number of clusters, $\mathbf{x}$ is an observation for each country assigned to cluster $C_q$, and $\bm{\mu}_q$ is the centroid of cluster $C_q$. The final optimal number is identified at the ``elbow" point of the curve, where the rate of decrease in inertia slows significantly, indicating diminishing returns in clustering quality with additional clusters. Finally, the distance measure used to define similarity between data points is the Euclidean distance metric. It calculates the straight-line distance between two points in an $n$-dimensional space.

\subsubsection{Method 1: Clustering based on raw time trend series}
\begin{figure}[ht]
\centering
\includegraphics[width=1\textwidth]{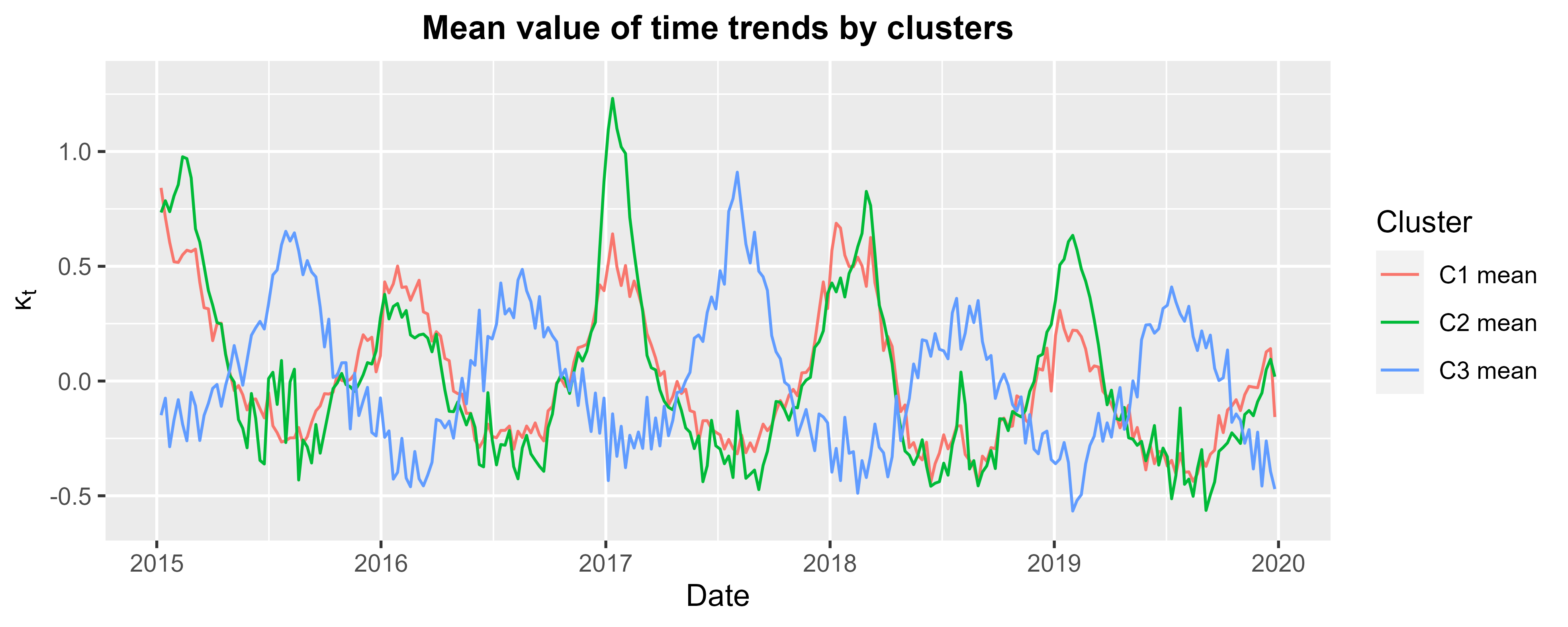}
\caption{Mean value of the time trends in each cluster}
\label{fig:M1: time trends by cluster}
\end{figure}

We begin by analysing the raw time series of the country-specific $\kappa_t$ as the data input. This approach identifies three distinct clusters, as detailed in Table \ref{tab:M1: country list}. Figure \ref{fig:M1: time trends by cluster} presents the mean value of the time trends for each cluster. Notably, Cluster 3, which comprises only two countries from the Southern Hemisphere, exhibits seasonal patterns that are opposite to those of the other clusters. Cluster 2 consists exclusively of European countries and displays more pronounced seasonal peaks relative to Cluster 1, highlighting its higher mortality rates in winter.

\begin{table}[ht]
\small{
  \centering
  \caption{List of countries in each cluster under Method 1}
    \vspace{0.3em}
    \begin{tabular}{ccp{28em}}
    \toprule
    \textbf{Cluster} & \textbf{Number of countries} & \multicolumn{1}{l}{\textbf{Country name}} \\
    \midrule
    1     & 11     & Canada, Denmark, England and Wales, Finland, Netherlands, Norway, Scotland, South Korea, Sweden, Taiwan, USA \\
    2     & 17    & Austria, Belgium, Bulgaria, Croatia, Czech Republic, France, Germany, Greece, Hungary, Italy, Lithuania, Poland, Portugal, Slovakia, Slovenia, Spain, Switzerland  \\
    3     & 2    & Australia, New Zealand \\
    \bottomrule
    \end{tabular}%
  \label{tab:M1: country list}%
  }
\end{table}%

It is noteworthy that Cluster 2 exhibits a pronounced spike during the winter of 2017, which can be attributed to an extreme cold-weather event, affecting most of Europe in January. Temperatures remain significantly below seasonal averages for several days, particularly impacting southern and southeastern Europe. Greece, for instance, records its highest January mortality rate in 2017 compared to any other January between 2013 and 2018, with individuals aged over 80 being most affected \citep{kostopoulou2023analysis}. Italy also experiences abnormal snowfall due to cold air moving across the relatively warmer Adriatic Sea, picking up moisture and generating intense precipitation \citep{murtas2019effects}. \cite{demirtacs2022anomalously} further investigates the broader effects of severe snowstorms and cold temperatures across southeastern Europe. A similar pattern is observed in 2019, with another cold wave, triggered by a Sudden Stratospheric Warming (SSW) event. It results in widespread snowfall and freezing temperatures across central and southern Europe, contributing to increased mortality and significant societal disruptions \citep{knight2021predictability}.

\subsubsection{Method 2: Clustering based on estimated trend slopes}
\begin{figure}[ht]
\centering
\includegraphics[width=1\textwidth]{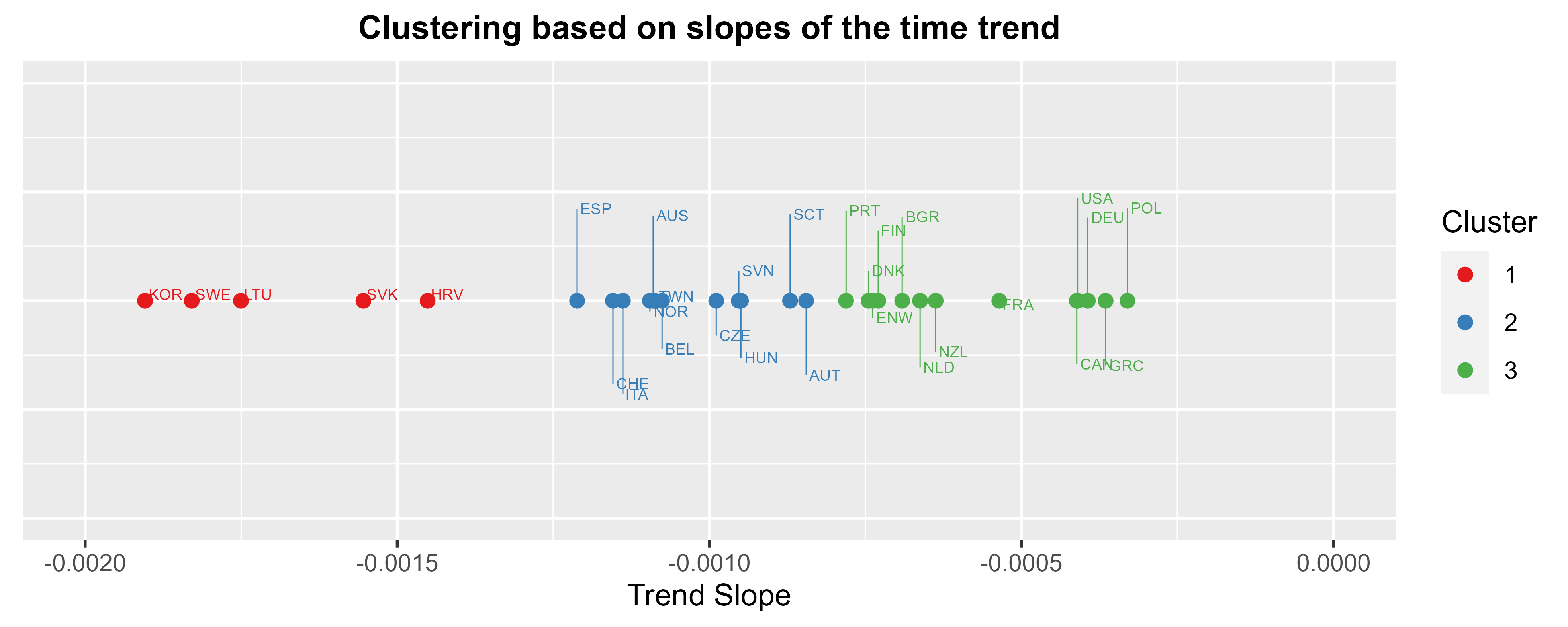}
\caption{Trend slopes by clusters}
\label{fig:Method_2_clustering}
\end{figure}

Building on the previous analysis of the raw time series, we now focus on the trend reflecting mortality improvement over the years. We first apply Seasonal and Trend decomposition using LOESS (STL) to the country-specific $\kappa_t$ series \citep{cleveland1990stl}. This method uses locally estimated scatterplot smoothing (LOESS) to decompose each time series into three distinct components: a long-term trend, a recurring seasonal pattern, and a residual representing irregular fluctuations. To classify countries based on their trend dynamics, we compute the slope of each decomposed long-term trend using linear regression. This procedure identifies three distinct clusters, as detailed in Table \ref{tab:M2 country list}. Figure \ref{fig:Method_2_clustering} displays the distribution of trend slopes by clusters. Countries in Cluster 1 exhibit the most substantial mortality improvement, whereas Cluster 2 countries have moderate improvement and Cluster 3 countries are more stable over the five-year period.

\begin{table}[ht]
\small{
  \centering
  \caption{List of countries in each cluster under Method 2}
    \vspace{0.3em}
    \begin{tabular}{ccp{28em}}
    \toprule
    \textbf{Cluster} & \textbf{Number of countries} & \multicolumn{1}{l}{\textbf{Country name}} \\
    \midrule
    1     & 5     & \multicolumn{1}{l}{Croatia, Lithuania, Slovakia, South Korea, Sweden} \\
    2     & 12    & Australia, Austria, Belgium, Czech Republic, Hungary, Italy, \newline{}Norway, Scotland, Slovenia, Spain, Switzerland, Taiwan   \\
    3     & 13    & Bulgaria, Canada, Denmark, England and Wales, Finland, \newline{}France, Germany, Greece, Netherlands, New Zealand, Poland, Portugal, USA \\
    \bottomrule
    \end{tabular}%
  \label{tab:M2 country list}%
  }
\end{table}%

All countries have negative trend slopes, indicating the mortality improvement across five years. The rapid mortality improvement observed in Cluster 1 may be attributed to factors such as healthcare system reforms and advances in medical technology and treatment \citep[see e.g.,][]{european2021state, pekarcikova2024cancer}. Given that countries like Croatia, Lithuania, and Slovakia have historically exhibited relatively higher mortality rates, these advancements suggest a particularly substantial improvement in health outcomes for these populations, resulting in higher trend slopes in absolute values. In Clusters 2 and 3, some geographical patterns are evident, including the grouping of Canada and the US, as well as several neighbouring European countries. These similarities likely reflect shared characteristics such as public health infrastructure, climatic conditions, and socio-economic contexts, which may contribute to comparable rates of mortality improvement over the period from 2015 to 2019.

\subsubsection{Method 3: Clustering based on estimated trend slopes and seasonal strength}
\begin{figure}[h!]
\centering
\includegraphics[width=1\textwidth]{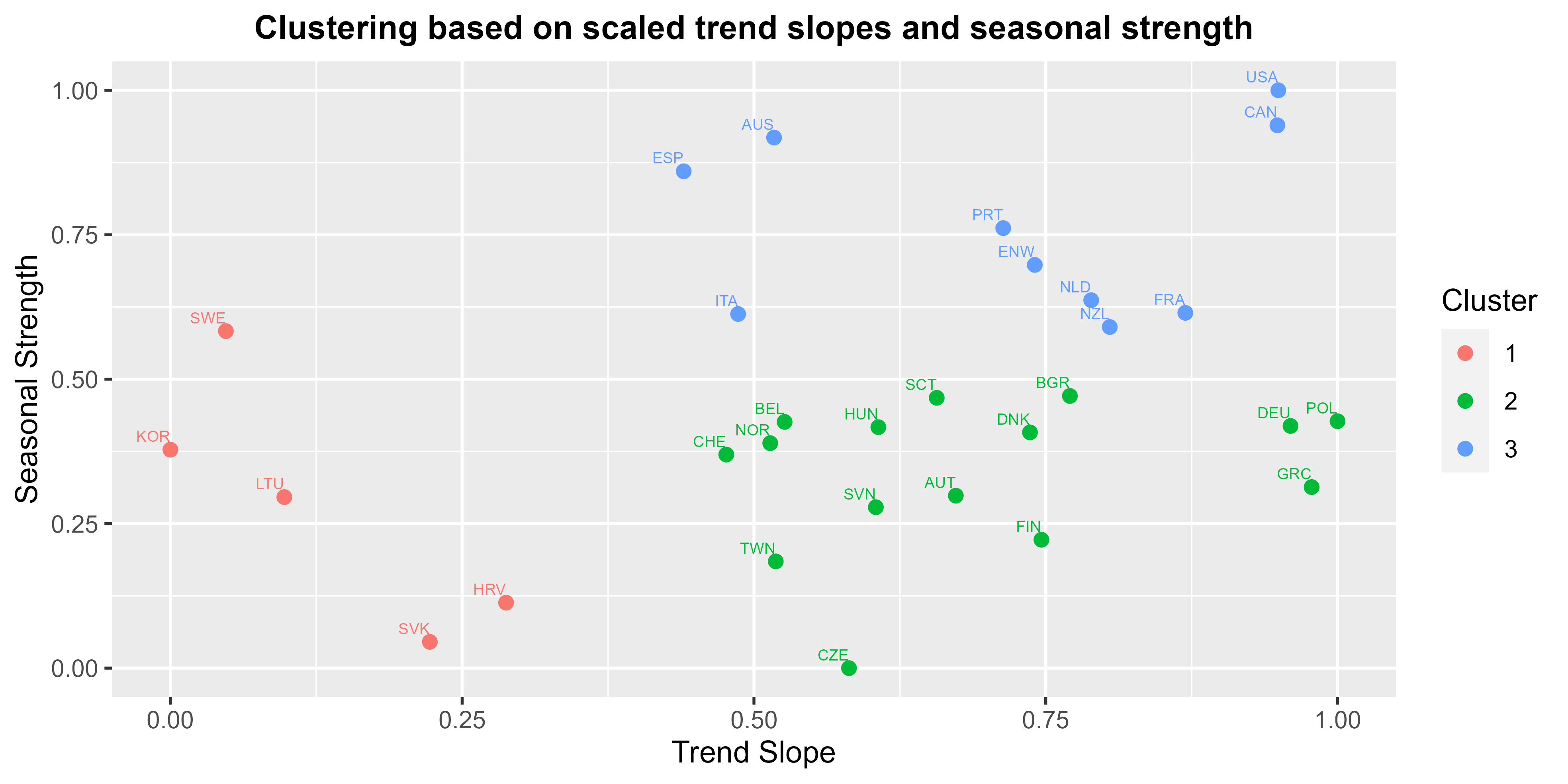}
\vspace{-0.05in}
\caption{Clusters based on min-max scaled trend slopes and seasonal strength}
\label{fig:Method_3_clustering}
\end{figure}

Finally, building upon the trend slopes employed in Method 2, we additionally incorporate the measure of seasonal strength derived from the aforementioned STL decomposition. These two features, trend and seasonality, are essential for characterising the underlying dynamics of mortality evolution. To ensure comparability between the two variables, min-max scaling is applied. The resulting three clusters are detailed in Table \ref{tab:M3 country list}. Figure \ref{fig:Method_3_clustering} presents a scatterplot, illustrating the clustering outcome. Cluster 1 is mainly determined by the trend slopes, where the distinction between Cluster 2 and Cluster 3 is primarily driven by differences in seasonal strength. Cluster 1 comprises countries exhibiting a higher rate of mortality improvement, as indicated by more negative trend slopes. Cluster 2 includes countries with relatively stable mortality patterns and weaker seasonality. Cluster 3 consists of countries with similar rates of mortality improvement to Cluster 2, but exhibiting stronger seasonal variations.

\begin{table}[ht]
\small{
  \centering
  \caption{List of countries in each cluster under Method 3}
    \vspace{0.3em}
    \begin{tabular}{ccp{28em}}
    \toprule
    \textbf{Cluster} & \textbf{Number of countries} & \multicolumn{1}{l}{\textbf{Country name}} \\
    \midrule
    1     & 5     & \multicolumn{1}{l}{Croatia, Lithuania, Slovakia, South Korea, Sweden} \\
    2     & 15    & Austria, Belgium, Bulgaria, Czech Republic, Denmark, Finland, Germany, Greece, Hungary, Norway, Poland, Scotland, Slovenia, Switzerland, Taiwan   \\
    3     & 10    & Australia, Canada, England and Wales, France, Italy, \newline{}Netherlands, New Zealand, Portugal, Spain, USA \\
    \bottomrule
    \end{tabular}%
  \label{tab:M3 country list}%
  }
\end{table}%

We investigate the countries in Cluster 3 with respect to their pronounced seasonal mortality patterns. These stronger seasonal fluctuations are potentially attributable to cold-weather effects, which contribute to elevated mortality rates during winter months. Furthermore, all countries in Cluster 3 have aging populations, wherein older adults exhibit increased vulnerability to seasonal health variations \citep{bonnet2021population}. \cite{falagas2009seasonality} analyse monthly mortality data to examine seasonal patterns across Mediterranean countries, including France, Italy, Spain, and Portugal, as well as Australia, Canada, and the US. Their findings indicate that the lowest average daily mortality occurs in September for countries in the Northern Hemisphere and in March for Australia and New Zealand. These more pronounced seasonal variations are further attributed to environmental factors, such as temperature and daylight duration.

\subsection{Forecast performance with expanding windows}
To mitigate randomness in the sampling procedure, we again employ a 10-fold expanding window approach. Table \ref{tab:forecast MAPE clustering} summarises the forecast performance across the clustering methods described above in Section \ref{sec:3_clustering_methods}. Among the three clustering approaches, Method 1, which uses raw time trend series, yields the poorest performance, likely due to its higher dimensionality and associated noise. In contrast, Method 2, which relies solely on the slopes of country-specific time trends, achieves the best forecast accuracy. This result suggests that utilising a more distinct and parsimonious clustering criterion may facilitate the grouping of structurally similar countries, thereby enhancing predictive performance. In this experiment, regardless of the clustering strategy employed, based on the available data, our proposed GBLL model outperforms both the LL and HBY models, with the latter exhibiting the weakest forecast accuracy.

\begin{table}[ht]
  \centering
  \caption{Mean MAPE of out-of-sample forecasts across 30 countries with clustering ($\times 100$ scale)}
    \begin{tabular}{ccccccccccc}
    \toprule
          & \multicolumn{3}{c}{\textbf{Method 1}} & \multicolumn{3}{c}{\textbf{Method 2}} & \multicolumn{3}{c}{\textbf{Method 3}} & \multicolumn{1}{c}{\textbf{Whole}} \\
    \midrule
    \makebox[0.5cm]{$h$} & \makebox[1cm]{LL} & \makebox[1.1cm]{HBY} & \makebox[1.2cm]{GBLL} & \makebox[1cm]{LL} & \makebox[1.1cm]{HBY} & \makebox[1.3cm]{GBLL} & \makebox[1cm]{LL} & \makebox[1.1cm]{HBY} & \makebox[1.3cm]{GBLL} & \makebox[1.3cm]{GBLL}\\
    \midrule
    1     & 5.459 & 6.246 & \textbf{5.309} & 5.357 & 5.924 & \textbf{5.199} & 5.453 & 5.867 & \textbf{5.299} & 5.278\\
    2     & 5.734 & 6.458 & \textbf{5.586} & 5.647 & 6.220 & \textbf{5.481} & 5.746 & 6.057 & \textbf{5.577} & 5.620\\
    3     & 5.774 & 6.476 & \textbf{5.625} & 5.715 & 6.242 & \textbf{5.540} & 5.783 & 6.106 & \textbf{5.605} & 5.582\\
    4     & 5.818 & 6.504 & \textbf{5.658} & 5.766 & 6.251 & \textbf{5.588} & 5.841 & 6.136 & \textbf{5.659} & 5.602\\
    5     & 5.908 & 6.500 & \textbf{5.741} & 5.816 & 6.322 & \textbf{5.632} & 5.902 & 6.151 & \textbf{5.713} & 5.671\\
    6     & 5.954 & 6.553 & \textbf{5.785} & 5.861 & 6.401 & \textbf{5.680} & 5.940 & 6.176 & \textbf{5.747} & 5.718\\
    7     & 5.997 & 6.576 & \textbf{5.828} & 5.882 & 6.455 & \textbf{5.703} & 5.963 & 6.183 & \textbf{5.772} & 5.753\\
    8     & 6.050 & 6.640 & \textbf{5.879} & 5.924 & 6.494 & \textbf{5.747} & 6.021 & 6.230 & \textbf{5.832} & 5.829\\
    9     & 6.091 & 6.689 & \textbf{5.917} & 5.955 & 6.522 & \textbf{5.779} & 6.061 & 6.281 & \textbf{5.875} & 5.878\\
    10    & 6.112 & 6.708 & \textbf{5.935} & 5.968 & 6.536 & \textbf{5.790} & 6.077 & 6.308 & \textbf{5.887} & 5.914\\
    11    & 6.110 & 6.689 & \textbf{5.924} & 5.959 & 6.503 & \textbf{5.777} & 6.072 & 6.300 & \textbf{5.876} & 5.922\\
    12    & 6.104 & 6.665 & \textbf{5.913} & 5.962 & 6.493 & \textbf{5.778} & 6.077 & 6.311 & \textbf{5.875} & 5.911\\
    \bottomrule
    \end{tabular}%
  \label{tab:forecast MAPE clustering}%
\end{table}%

To further assess the impact of clustering, we compare the results with the case without clustering, as presented in Table \ref{tab:forecast_MAPE_whole}. The overall performance remains comparable, however, Method 2 is the only clustering strategy that improves forecast accuracy across all models and forecast horizons relative to the baseline case without clustering. The varying performance may be attributed to the heterogeneity in the rate of mortality improvement and the degree of seasonality across individual countries, particularly given that weekly mortality data tend to exhibit higher-frequency fluctuations and varying levels of noise. These findings suggest that, within this dataset of 30 countries, trend slopes serve as the most effective factor for grouping countries with similar mortality experience together. Additionally, we include a reference column reporting the forecast MAPE of the GBLL model without any clustering. Notably, this unclustered version of the GBLL model still outperforms all variants of the LL and HBY models under each of the three clustering approaches and across all forecast horizons. Based on the available data, the empirical evidence highlights that the GBLL framework achieves superior forecast performance without the need for clustering, surpassing baseline models that rely on clustering to enhance coherence and forecast accuracy.

Focusing on the HBY model, which emphasises the need for grouping countries with similar characteristics to enhance forecast performance, the results lend support to this premise. All three clustering methods improve forecast accuracy compared to the unclustered baseline, with the exception of the first month under Method 1. However, identifying an optimal clustering strategy that yields substantial improvement on all models over the baseline case without clustering remains challenging. This may be attributed to the limited span of data with only five years, during which there is minimal divergence in mortality trends, resulting in relatively homogeneous raw time trend series across countries. These observations underscore the importance of a modelling framework that is both adaptive and insensitive, regardless of clustering choices. The LL model demonstrates relatively similar performance across various clustering methods and may be regarded as a stable, though less flexible, approach. However, this stability comes at the cost of lower predictive accuracy. Conversely, the HBY model exhibits substantial sensitivity to country groupings and continues to deliver the weakest forecast performance overall. In contrast, the proposed GBLL model distinguishes itself by maintaining high forecast accuracy across all clustering strategies. Its performance is largely unaffected by the social, economic, or environmental heterogeneity of the countries, highlighting its insensitivity and superior forecasting capability.

\section{Conclusion}
\label{sec:conclusion}
This paper proposes a multi-population mortality model, the gradient boosted Li and Lee model that integrates gradient boosting techniques to iteratively capture the heterogeneous seasonal patterns among all age groups and countries. Specifically, we apply this model to weekly mortality rates across 30 countries for the period 2015--2019. Empirical findings demonstrate that the GBLL model outperforms the benchmark models, including the original Li and Lee model and the Hyndman--Booth--Yasmeen model of order six, in both in-sample fitting and out-of-sample forecasts. In addition, we conduct an experiment on clustering to group countries with similar mortality dynamics. Empirical results indicate that the performance varies depending on the clustering method employed, and clustering does not universally lead to better outcomes. Nevertheless, the proposed GBLL model exhibits insensitivity to the choice of country groupings, maintaining stable and superior forecast accuracy across different clustering arrangements. This adaptability provides evidence of the flexibility of the GBLL model in handling heterogeneous data without the need for extensive preprocessing or manual adjustment.

The proposed model provides several promising directions for future research. First, the reciprocal transformation of mortality rates for Southern Hemisphere countries may be leveraged in mortality risk hedging strategies, given the inverse seasonal patterns and the offset in the timing of mortality peaks and troughs across hemispheres. Second, the gradient boosting framework may be extended to other stochastic mortality models, such as the Cairns--Blake--Dowd model \citep{cairns2006two} and its variants. Further extensions could incorporate finer age stratification, such as 5-year or 10-year age intervals, as well as additional dimensions, including sex, causes of death, or regional subgroups, to provide more insights into granular dynamics and improve the precision of mortality modelling. However, this would depend on the availability of such detailed data. Finally, the proposed framework is well suited to within-year mortality forecasting, enabling early-season assessments of expected winter mortality and facilitating timely monitoring of unusually elevated mortality levels linked to extreme events, including influenza seasons and pandemics.

\section*{Acknowledegments}  
The authors thank Rob J. Hyndman of Monash University, Australia, Li Li of the University of Science and Technology Beijing, China, and Hanlin Shang of Macquarie University, Australia for their helpful comments on an earlier version of this paper. Ziting Miao is supported by the Australian Government Research Training Program Scholarship and the Henry Buck Scholarship from the University of Melbourne.

\section*{Competing interest} 
The authors declare no conflicts of interest or competing interests in this paper.

\section*{Appendices}
\begin{appendices}
\section{Mortality data for 30 countries}
\label{sec:Appendix_A} 
While Figure \ref{fig:30_country_data} does not clearly reveal country-specific mortality trends, the overarching patterns discussed in Section \ref{sec:data_visualisation} remain similar.

\setcounter{figure}{0}
\renewcommand{\thefigure}{\thesection.\arabic{figure}}
\renewcommand{\theHfigure}{\thesection.\arabic{figure}}
\begin{figure}[h!]
\centering
\includegraphics[width=1\textwidth]{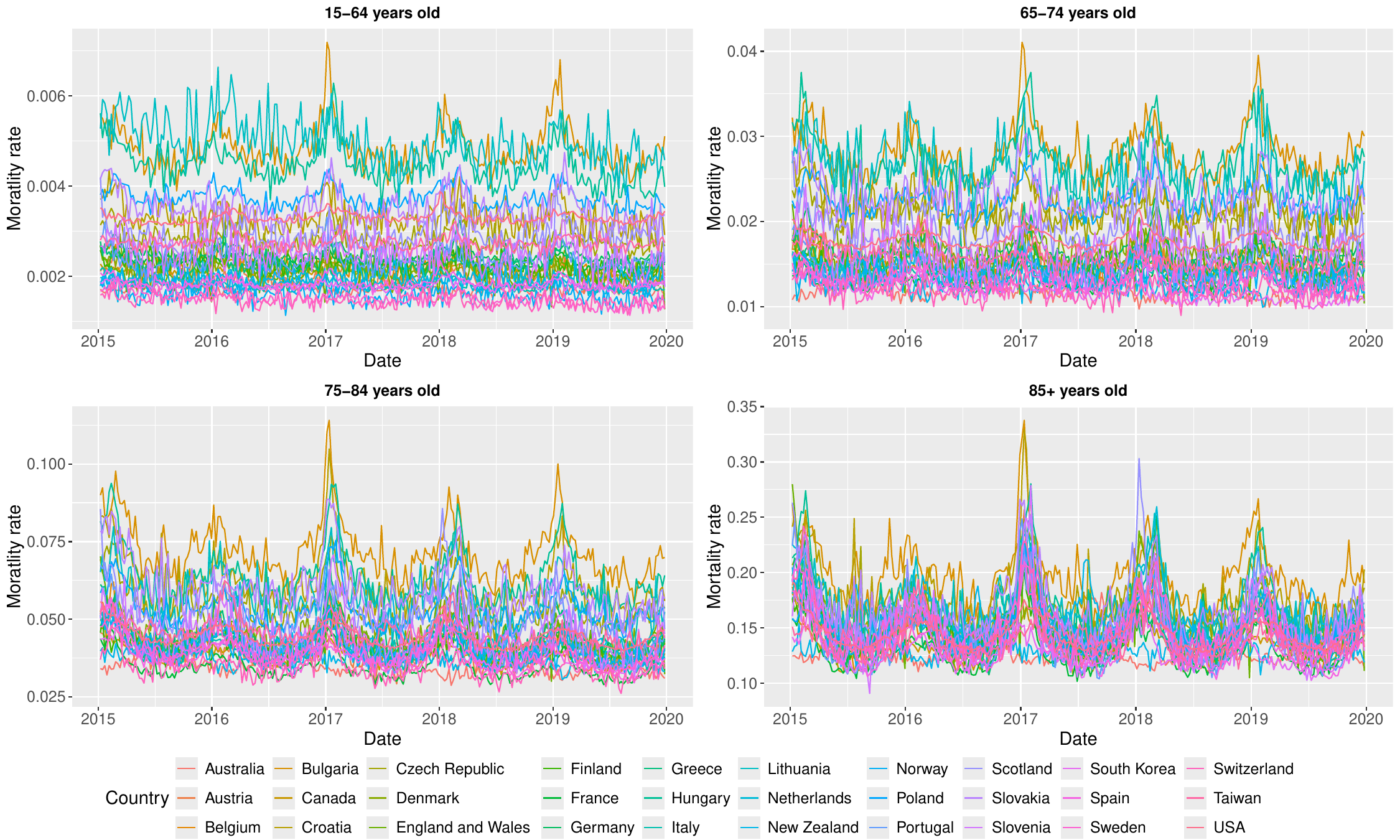}
\caption{2015--2019 mortality rate for 30 countries by age groups}
\label{fig:30_country_data}
\end{figure}

\section{The Lee--Carter model}
\label{sec:Appendix_B}
The Lee--Carter model \citep{lee1992modeling} is the first single-population stochastic mortality model and  serves as the foundation for the estimation of the Li and Lee model. This model separates the historical mortality rate into age-specific and time-varying components. For age group $x\in\{1,...,N\}$ and time $t\in\{1,...,T\}$, the method is formulated as follows:
\begin{equation*}
\log(m_{x,t})=a_x + b_x \kappa_t + \varepsilon_{x,t},
\end{equation*}
where $m_{x,t}$ is the mortality rate for age group $x$ at time $t$, $a_x$ represents the age-specific mean of log mortality rate over time, $\kappa_t$ is the mortality improvement over time shared by all age groups,  $b_x$ is the corresponding age-specific loading, and $\varepsilon_{x,t}$ is an error term. The model contains two constraints to ensure the uniqueness of parameter estimates:
\begin{equation*}
    \sum_{x=1}^N{b_x}=1 \mbox{~~and~~} \sum_{t=1}^T{\kappa_t}=0.
\end{equation*}

The estimation of the Lee--Carter model is via Singular Value Decomposition (SVD). First let $\bm{Y}$ denote a $T \times N$ (time $\times$ age group) matrix with $\log(\tilde{m}_{x,t})$ in row $t$ and column $x$. The age-specific intercept $\hat{a}_x$ is calculated by taking the sample mean of the logarithmic mortality rate for each age group $x$ over $T$ periods. Then, subtract the age-specific intercept from each row of $\bm{Y}$ and denote the resulting matrix as $\bm{\bar{Y}}$. The singular value decomposition of $\bm{\bar{Y}}$ gives the estimated age-specific coefficients $\hat{b}_x$ and the estimated time component $\hat{\kappa}_t$. The completed version of the estimation procedure is summarised as follows in Algorithm \ref{arg:LC}. A wide range of extensions of the Lee--Carter model have been developed and published \citep[see e.g.,][]{brouhns2002poisson, cairns2006two, renshaw2006cohort, hyndman2007robust}.

\renewcommand{\thealgocf}{B.\arabic{algocf}}
\setcounter{algocf}{0}
\begin{algorithm}[ht]
\setstretch{1.6}
\caption{Estimation of the Lee--Carter model}
\label{arg:LC}
\KwInput{$\bm{Y}$ (a $T \times N$ matrix)}

\KwOutput{$\hat{y}_{x,t}$ for $x\in\{1,...,N\}$, $t\in\{1,...,T\}$, $\bm{\hat{\kappa}}_t=(\hat{\kappa}_1,\hat{\kappa}_2,...,\hat{\kappa}_T), \bm{\hat{a}}_x=(\hat{a}_1,\hat{a}_2,...,\hat{a}_N)$ \textbf{and} $\bm{\hat{b}}_x=(\hat{b}_1,\hat{b}_2,...,\hat{b}_N)$}

\KwProcedure{LC($\bm{Y}$)}
\begin{algorithmic}[1]
\State $\hat{a}_x \gets \frac{1}{T} \sum_{t=1}^{T} \log(\tilde{m}_{x,t}), \forall x$

\State $\bar{y}_{x,t} \gets \log(\tilde{m}_{x,t})-\hat{a}_x$

\State Form $\bm{\bar{Y}}=(\bar{\bm{y}}_{1,t}, \bar{\bm{y}}_{2,t},...,\bar{\bm{y}}_{N,t})$

\State Do the Singular Value Decomposition (SVD) on $\bm{\bar{Y}}$

\State $\bm{\hat{\kappa}}_t \gets d\times \bm{u}\times \sum_{x=1}^{N}{v_x}$, where $d$ is the first singular value of $\bm{\bar{Y}}$, $\bm{u}$ is the first column of \hspace{2em} the left singular vectors of $\bm{\bar{Y}}$ and $\bm{v}$ is the first column of the right singular vectors of $\bm{\bar{Y}}$

\State $\bm{\hat{b}}_x \gets \frac{\bm{v}}{\sum_{x=1}^{N}{v_x}}, \forall x$

\State $\hat{y}_{x,t}=\log(\hat{m}_{x,t}) \gets \hat{a}_x+\hat{b}_x \hat{\kappa}_t$, $\forall x,t$
\end{algorithmic}
\end{algorithm}

\section{The Hyndman--Booth--Yasmeen (HBY) model}
\label{sec:HBY}
\cite{hyndman2013coherent} propose a generalised Li and Lee model using the product-ratio method and apply Principal Component Analysis to both product and ratio functions to forecast sub-population mortality coherently. The aggregated model for each country $j\in\{1,...,J\}$ is given by: For age group $x\in\{1,...,N\}$ and time $t\in\{1,...,T\}$,
\begin{equation*}
\log(m^{j}_{x,t})=\mu_{x,j} + \sum_{r=1}^{R} \beta_{t,r}  \phi_{x,r} + \sum_{u=1}^{U} \gamma^j_{t,u}  \psi^j_{x,u} + z^j_{x,t},
\end{equation*}
where $\mu_{x,j}$ is the mean of log mortality rate for age group $x$ and country $j$ over time, and $\{\phi_{x,r}\}$ and $\{\psi^j_{x,u}\}$ are two sets of orthonormal basis functions for age group $x$, country $j$, index $r\in\{1,...,R\}$ and $u\in\{1,...,U\}$. In addition, $\beta_{t,r}$ represents the common time trend and $\gamma^j_{t,u}$ is the country-specific time trend for time $t$, country $j$, index $r\in\{1,...,R\}$ and $u\in\{1,...,U\}$, respectively. Finally, $z^j_{x,t}$ is the error term. The HBY model can be implemented via the R package \texttt{demography}. Following \cite{hyndman2013coherent}, we employ the HBY model with order 6 as a baseline for comparative performance evaluation. While determining the optimal order of the HBY model for high-frequency mortality data is of importance, we leave it for future investigation.

\section{The Vector Auto-Regressive models}
\label{sec:VAR}
The Vector Auto-Regressive (VAR) model is one of the most flexible and easy to use models for the analysis of multivariate time series \citep{sims1980macroeconomics}. It treats all series of the weekly mortality data jointly endogenous, allowing each age group or population to depend on its own lags and those of others. This structure captures both temporal dynamics and cross-group dependencies, which is crucial for multi-population mortality forecasting. However, the VAR framework assumes stationarity, requiring appropriate data preprocessing including normal and seasonal differencing, given the noisy weekly mortality rate with annual cyclical patterns.

First, we consider the single-population VAR model. Based on the observation from Figure \ref{fig:4_country_data} and Figure \ref{fig:30_country_data}, the mortality rate series are typically non-stationary. Therefore, we first take the seasonal difference for each country $j$ if necessary, denoted by $z^j_{x,t}=y^j_{x,t} - y^j_{x,t-52}$. If the differenced series is still not stationary, we will follow \cite{guibert2019forecasting} to take the first difference of $z^j_{x,t}$, denoted by  $v^j_{x,t}=z^j_{x,t} - z^j_{x,t-1}$ to remove the linear trend. The series represents the log-mortality improvement rates. The VAR$(p)$ model for $\mathbf{v}_t^j=(v^j_{1,t},...,v^j_{N,t})$ is given by: for country $j \in\{1,...,J\},$
\begin{equation*}
    \mathbf{v}_t^j = \mathbf{c}^j + \sum_{i=1}^{p} A_i^j \mathbf{v}_{t-i}^j + \boldsymbol{\varepsilon}_t^j,
\end{equation*}
where $\mathbf{c}^j \in \mathbb{R}^x$ is a vector of intercept terms, $p$ is the order, $A_i^j \in \mathbb{R}^{x \times x}$ for $i=1,..,p$, are autoregressive coefficient matrices and $\boldsymbol{\varepsilon}_t^j$ is the error term. The lag order $p$ of the VAR model is selected using the Hannan--Quinn (HQ) information criterion, which balances goodness-of-fit and model parsimony through a penalty term proportional to $2k\log(\log N)$, where $k$ is the total number estimated parameters and $N$ is the total number of observations. The HQ criterion balances the overfitting nature of the Akaike Information Criterion (AIC) and the underfitting of the Bayesian Information Criterion (BIC)\citep[see e.g.,][]{hatemij2003new, ivanov2005note}. The VAR model can be implemented via the R package \texttt{vars} and the out-of-sample forecast is carried out via the standard \texttt{predict} function.

The Global Vector Auto-Regressive (GVAR) model \citep{pesaran2004gvar} extends the standard VAR frameworks by jointly modelling multiple interconnected time series through country-specific VARX models augmented with weighted foreign variables. Cross-country dependence is captured via these foreign aggregates rather than through a fully unrestricted multivariate system, allowing for scalable estimation and coherent global dynamics. The GVAR model consists of country-specific VARX models of the form:
\begin{equation*}
\mathbf{v}_t^j = \mathbf{c}^j + \sum_{p=1}^{P^j} A_p^j\mathbf{v}_{t-p}^j + \sum_{q=0}^{Q^j} B_{q}^j\mathbf{v}_{t-q}^{j*} + \boldsymbol{\varepsilon}_t^j,
\end{equation*}
where $B_q^j \in \mathbb{R}^{x \times x}$ for $q=1,..,Q^j$, are autoregressive coefficient matrices for the foreign variables $\mathbf{v}_{t-q}^{j*}$. The foreign variables $\mathbf{v}_{t}^{j*}$ entering the country-specific models are defined as weighted averages of other countries’ variables:
\begin{equation*}
\mathbf{v}_{t}^{j*} = \sum_{j \neq i} w_{ij} \mathbf{v}_{t}^i,
\end{equation*}
where $w_{ij}=1/29$ for $j\neq i$ and $w_{ij}=0$ otherwise, since we have 30 countries in the modelling process. The GVAR model can be implemented via the R package \texttt{GVARX}. However, the order can only be chosen between 1 and 2. The out-of-sample forecast is carried out following the approach detailed in Section 2.2 of \cite{di2013gvar}.

\section{The country-specific trends under the GBLL model}
\label{sec:Appendix_C}
In addition to the common trends discussed in Section \ref{sec:fitting}, we present a detailed decomposition of the country-specific components under the GBLL model for the first two iterations in Figure \ref{fig:GBLL_components_country}.

\begin{itemize}
    \item For the age-specific intercepts $\hat{a}_x$, Australia and New Zealand exhibit an opposite pattern relative to the Northern Hemisphere countries. This is due to the use of the reciprocal transformation of mortality rates. As with the common components, the primary effects are largely captured in the first iteration, as evidenced by the larger magnitudes.

    \item For the age loadings $\hat{b}_x$ in the first iteration, Slovenia displays elevated mortality sensitivity among individuals aged 65--74. Denmark shows comparatively higher exposure in the 15--64 age group, despite having relatively favourable outcomes for the 65--74 cohort compared to most other countries. Additionally, higher exposure is observed across all age groups except for those aged 85 and above in Netherlands. In the second iteration, the country-specific components reveal particularly pronounced deviations for Belgium and Finland, suggesting substantial excess loadings relative to other countries.

    \item For the time-varying index $\hat{\kappa}_t$, the clear W-shaped seasonal pattern evident in the common trend is often obscured at the individual country level, largely due to local noise and country-specific variability. The first iteration captures the majority of temporal dynamics, as indicated by the larger magnitudes of the components and a lower degree of white-noise behaviour relative to the second iteration. Nonetheless, several countries exhibit unexpected behaviours for the second iteration. For example, Denmark shows a sharp downward spike in January 2017, while Netherlands experiences a pronounced upward spike in March 2018.
\end{itemize}
\vspace{-0.1in}
\renewcommand{\thefigure}{E.\arabic{figure}}
\setcounter{figure}{0}
\begin{figure}[htbp]
\centering
\includegraphics[width=0.95\textwidth]{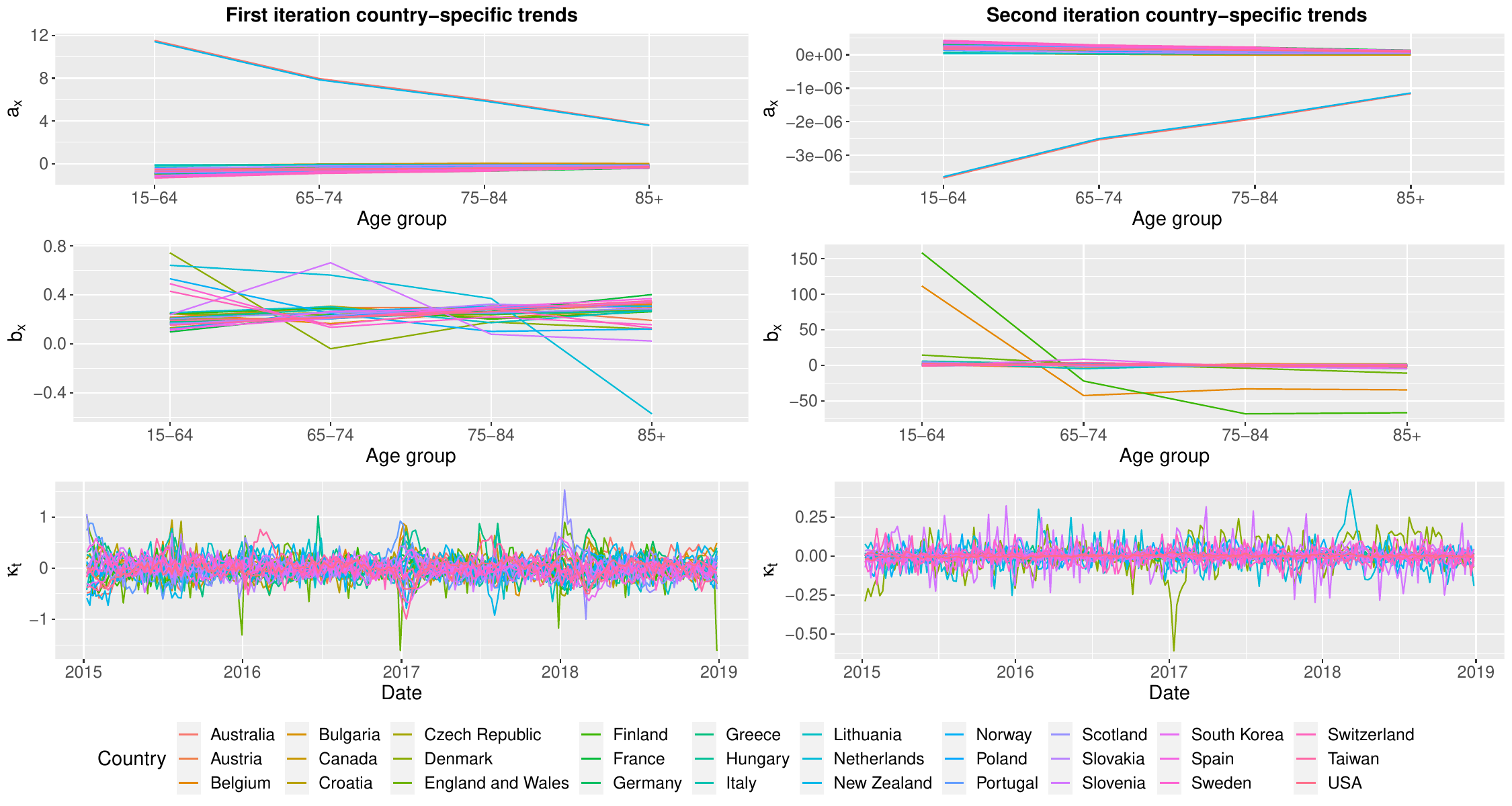}
\vspace{-0.15in}
\caption{Components of the country-specific trends under the GBLL model: first iteration (left) and second iteration (right)}
\label{fig:GBLL_components_country}
\end{figure}

\end{appendices}

\clearpage

\SetKwInput{KwInput}{Input} 
\SetKwInput{KwOutput}{Output} 
\SetKwInput{KwProcedure}{Procedure}

\setlength{\bibsep}{1pt}
\topmargin -1.30cm \oddsidemargin -0.00cm \evensidemargin 0.0cm
\textwidth 16.56cm \textheight 23.20cm

\parindent 5ex

\bibliographystyle{apalike}
\begin{center}
{\LARGE \textbf{Supplementary Material}}\\[3em]
\end{center}

\appendix
\numberwithin{table}{section}
\numberwithin{figure}{section}

\section{Alternative parameter modelling choices of the GBLL model}
To demonstrate the insensitivity of the GBLL framework, we have substantially expanded the methodological discussion in Section 3.2 to examine alternative choices of loss functions, stopping criteria, models used for forecasting time trends, and seasonal adjustments for countries in the Southern Hemisphere. The main additions are summarised below. Each modification is implemented separately, while all other components of the framework are held fixed.

\subsection{Loss function}
To evaluate the sensitivity of our results to the choice of loss function, we additionally consider the absolute (L1) loss function alongside the quadratic (L2) loss function employed in the manuscript. The L2 loss penalises larger errors more heavily and is optimal when the errors are approximately Gaussian, whereas the L1 loss treats all deviations equally and is more robust to outliers or extreme values. The absolute loss function is defined as:
    \begin{equation*}
     L_g(\gamma)=
        \begin{cases}
        \sum_{j=1}^{J}\sum_{x=1}^{N}|\bm{y}_{x}^j-\gamma \bm{\hat{y}}_{x,1}^j|, & \text{if } g = 1, \\
        \sum_{j=1}^{J}\sum_{x=1}^{N}|\bm{e}_{x,g-1}^j - \gamma \bm{\hat{y}}_{x,g}^j|, & \text{if } g \geq 2,
        \end{cases}
    \end{equation*}
    where $\bm{y}_{x}^j$ denotes the log mortality rate for country $j$ and age group $x$, $\gamma_g$ is the learning rate associated with the $g$-th iteration in the gradient boosting algorithm, $\bm{\hat{y}}_{x,g}^j$ represents the fitted log mortality rate from the Li and Lee model for country $j$, age group $x$, at the $g$-th iteration, and $\bm{e}_{x,g}^j$ is the corresponding residual. The results in Table \ref{tab:R1_choices} indicate that, for our high-frequency weekly mortality data, forecasts obtained under both loss functions are nearly identical, demonstrating the insensitivity of the GBLL model to the choice of loss function. A likely explanation comes from the well-behaved residual structure of weekly mortality data after accounting for common and country-specific seasonal components, with extreme outliers being limited due to the incremental error correction inherent in the boosting framework.

    \begin{table}[htbp]
  \centering
  \caption{Mean MAPE of out-of-sample forecasts across 30 countries ($\times 100$ scale) for the GBLL model on different choices of the components}
    \begin{tabular}{cccccc}
    \toprule
    \makebox[0.6cm]{$h$}     & \makebox[1.8cm]{\textbf{Original}} & \makebox[2.8cm]{\textbf{Absolute loss}} & \makebox[3.1cm]{\textbf{Stopping criterion}} & \makebox[2.3cm]{\textbf{SARIMA}} & \makebox[3.4cm]{\textbf{Seasonal adjustment}} \\
    \midrule
    1     & 5.278 & 5.279 & 5.278 & 5.621 & 5.262 \\
    2     & 5.620 & 5.621 & 5.620 & 6.085 & 5.640 \\
    3     & 5.582 & 5.582 & 5.582 & 6.243 & 5.597 \\
    4     & 5.602 & 5.602 & 5.602 & 6.277 & 5.610 \\
    5     & 5.671 & 5.672 & 5.671 & 6.247 & 5.685 \\
    6     & 5.718 & 5.719 & 5.718 & 6.214 & 5.720 \\
    7     & 5.753 & 5.754 & 5.753 & 6.212 & 5.745 \\
    8     & 5.829 & 5.829 & 5.829 & 6.297 & 5.811 \\
    9     & 5.878 & 5.879 & 5.878 & 6.366 & 5.852 \\
    10    & 5.914 & 5.914 & 5.914 & 6.409 & 5.875 \\
    11    & 5.922 & 5.922 & 5.922 & 6.419 & 5.877 \\
    12    & 5.911 & 5.911 & 5.911 & 6.420 & 5.866 \\
    \bottomrule
    \end{tabular}%
  \label{tab:R1_choices}%
\end{table}%

\subsection{Stopping criterion}
For the stopping criterion, we have implemented an alternative early stopping rule in which the boosting algorithm terminates once the absolute difference between two consecutive loss function values falls below a small positive number, here chosen to be $10^{-6}$, in addition to the white-noise residual requirement employed in the manuscript. This constitutes an early stopping mechanism, as the incremental reduction in residual error diminishes rapidly across iterations for weekly mortality data. As shown in Table \ref{tab:R1_choices}, forecast performance is essentially unchanged with or without early stopping. Since early stopping functions as a regularisation mechanism, these results provide empirical evidence that the GBLL framework does not suffer from harmful overfitting, and that additional boosting iterations primarily serve to correct residual underfitting in the benchmark models without worsening out-of-sample performance.

To account for the additional model complexity arising from the increased number of boosting iterations relative to the benchmark LL and HBY models, we also consider the AIC and BIC model selection criteria to identify the iteration that yields the best performance. This requires the assumption that the error term $\varepsilon^j_{x,t}$ in the Li and Lee model follows a Gaussian distribution with mean zero and variance $\sigma^2$ in 
\begin{equation*}
    \log(m^{j}_{x,t})=A^j_x + b^p_x \kappa^p_t + b^j_x \kappa^j_t + \varepsilon^j_{x,t}.
\end{equation*} This assumption is consistent with the existing literature \citep[see e.g.,][]{li2021mortality, li2021coherent, scognamiglio2024multi}. Let $y_{x,t}^j=\log(\tilde{m}^j_{x,t})$ and $\hat{y}_{x,t}^j=\log(\hat{m}^{j}_{x,t})= \hat{A}^j_x+\hat{b}^p_x \hat{\kappa}^p_t+\hat{b}^{j}_x \hat{\kappa}^{j}_t$, the log-likelihood across all country $j$, age group $x$ and time $t$ under this setting is: 
\begin{equation*}
    L_{\text{log-likelihood}}(\sigma^2) = -\frac{1}{2}\sum_{j,x,t}\left[\frac{\bigl(y_{x,t}^j - \hat{y}_{x,t}^j)^2}{\sigma^2} + \log(2\pi\sigma^2)\right].
\end{equation*} Let $N$ denote the total number of observations and define the residual sum of squares (RSS) as $\mathrm{RSS}= \sum_{j,x,t} \bigl(y_{x,t}^j - \hat{y}_{x,t}^j)^2,$ then the maximum likelihood estimator for $\sigma^2$ is $\hat{\sigma}^2=RSS/N$. Therefore, the maximum log-likelihood can be written as 
\begin{equation*}
    L_{\text{log-likelihood}}(\sigma^2) = -\frac{N}{2} \left[\log(2\pi) + 1 + \log\!\left(\frac{\mathrm{RSS}}{N}\right)\right].
\end{equation*}

The model selection criteria are based on the Akaike Information Criterion (AIC) and the Bayesian Information Criterion (BIC), both of which balance goodness-of-fit with model complexity. The Akaike Information Criterion (AIC) is defined as
\begin{equation*}
    \mathrm{AIC} = -2 L_{\text{log-likelihood}}(\sigma^2) + 2 k,
\end{equation*} where $k$ is the total number of estimated parameters. Additionally, the Bayesian Information Criterion (BIC) is given by
\begin{equation*}
    \mathrm{BIC} = -2 L_{\text{log-likelihood}}(\sigma^2) + k \log N.
\end{equation*}

The results indicate that the AIC is minimised when the number of iterations ranges from 48 to 50 for the 10-fold expanding window, which aligns with the configuration of our proposed GBLL model with a maximum of 50 iterations. In contrast, the BIC is minimised at a single iteration, corresponding to the Li and Lee model. This outcome is expected, as AIC prioritises predictive accuracy, whereas BIC emphasises model parsimony, imposing a stronger penalty on complexity and favoring simpler models \citep{burnham2002model}. Given that our primary objective is to provide a framework capable of delivering more accurate forecasts in a timely manner, model selection based on AIC supports the choice of the GBLL model despite its higher complexity.

\subsection{Time trend extrapolation}
With regard to the models used to extrapolate both the common and country-specific time trends $\kappa_t$, we additionally consider a Seasonal Auto-Regressive Integrated Moving Average (SARIMA) specification, in addition to the hybrid approach of Fourier regressors and ARIMA models proposed in the manuscript. A SARIMA model for a time series has orders $(p,d,q)\times(P,D,Q)_s$. The integers $p$, $d$, and $q$ denote the orders of the non-seasonal autoregressive (AR), differencing, and moving average (MA) components, respectively, while $P$, $D$, and $Q$ represent the corresponding orders for the seasonal components. The seasonal period $s$ controls the frequency at which seasonal effects repeat and is set to 52 to reflect annual seasonality in weekly data. However, in the weekly mortality setting considered here, SARIMA models require seasonal differencing at lag 52, which effectively removes one full year of observations from the estimation sample. Given that our analysis is based on five years of weekly data (260 observations), such seasonal differencing leads to a substantial loss of information and can result in unstable parameter estimations and deteriorated forecast performance.

\begin{table}[htbp]
  \centering
  \caption{Mean MAPE of out-of-sample forecasts across 30 countries ($\times 100$ scale) under alternative time trend forecasting models}
    \begin{tabular}{ccccccc}
    \toprule
          & \multicolumn{3}{c}{\textbf{Fourier-ARIMA}} & \multicolumn{3}{c}{\textbf{SARIMA}} \\
    \midrule
    \makebox[1.5cm]{$h$}     & \makebox[1.8cm]{\textbf{LL}}    & \makebox[2.1cm]{\textbf{HBY}}   & \makebox[2.1cm]{\textbf{GBLL}}  & \makebox[1.8cm]{\textbf{LL}}    & \makebox[2.1cm]{\textbf{HBY}}   & \makebox[2.1cm]{\textbf{GBLL}} \\
    \midrule
    1     & \textbf{5.436} & \textbf{5.985} & \textbf{5.278} & 5.747 & 6.589 & 5.621 \\
    2     & \textbf{5.779} & \textbf{6.512} & \textbf{5.620} & 6.191 & 7.573 & 6.085 \\
    3     & \textbf{5.746} & \textbf{6.537} & \textbf{5.582} & 6.345 & 7.838 & 6.243 \\
    4     & \textbf{5.774} & \textbf{6.547} & \textbf{5.602} & 6.380 & 7.707 & 6.277 \\
    5     & \textbf{5.846} & \textbf{6.553} & \textbf{5.671} & 6.351 & 7.436 & 6.247 \\
    6     & \textbf{5.889} & \textbf{6.563} & \textbf{5.718} & 6.324 & 7.294 & 6.214 \\
    7     & \textbf{5.923} & \textbf{6.610} & \textbf{5.753} & 6.324 & 7.264 & 6.212 \\
    8     & \textbf{5.998} & \textbf{6.698} & \textbf{5.829} & 6.415 & 7.333 & 6.297 \\
    9     & \textbf{6.048} & \textbf{6.752} & \textbf{5.878} & 6.486 & 7.429 & 6.366 \\
    10    & \textbf{6.085} & \textbf{6.758} & \textbf{5.914} & 6.534 & 7.447 & 6.409 \\
    11    & \textbf{6.096} & \textbf{6.745} & \textbf{5.922} & 6.552 & 7.446 & 6.419 \\
    12    & \textbf{6.091} & \textbf{6.726} & \textbf{5.911} & 6.563 & 7.453 & 6.420 \\
    \bottomrule
    \end{tabular}%
  \label{tab:SARIMA}%
\end{table}%

As shown in Table \ref{tab:R1_choices}, SARIMA-based extrapolation performs worse than the proposed hybrid approach, supporting our original modelling choice. Nonetheless, the resulting forecasts in Table \ref{tab:SARIMA} highlight that the GBLL model still outperforms both the LL and HBY models even when all time trends are fitted and forecasted using SARIMA models.

Moreover, from a computational perspective, SARIMA models with long seasonal periods are considerably more expensive to estimate and forecast, particularly when embedded within a multi-population, gradient boosting framework. The Fourier-ARIMA specification is substantially faster (around $1/18$ of the time required by the SARIMA approach) to fit and forecast, as it avoids the estimation of high-order seasonal autoregressive and moving average components, making it more suitable for our proposed GBLL model.

\subsection{Seasonal adjustment for Southern Hemisphere countries}
For the seasonal adjustment for Southern Hemisphere countries, we adopt an alternative symmetric transformation centred around historical mean mortality levels, in addition to the reciprocal transformation chosen in the manuscript. Specifically, we flip the peaks and troughs
in the observed mortality data for Southern Hemisphere countries. This transformation has the form: For country $j\in\{1,...,J\}$, age group $x\in\{1,...,N\}$ and time $t\in\{1,...,T\}$, $$\log(\tilde{m}_{x,t}^{j*})=\tilde{b}_x^j - \left (\log(\tilde{m}_{x,t}^j) - \tilde{b}_x^j\right),$$ where $\tilde{m}_{x,t}^{j*}$ and $\tilde{m}_{x,t}^j$ are the observed transformed and raw mortality rates for country $j$, age group $x$ and time $t$, respectively. $\tilde{b}_x^j$ denotes the average value of observed historical mortality rates for country $j$ and age group $x$ up to the current estimation window. For example, for the first expanding window with a training size of 169 weeks, $\tilde{b}_x^j$ is the mean observed mortality rates for country $j$ and age group $x$ over 169 weeks.

   \begin{table}[htbp]
  \centering
  \caption{Mean MAPE of out-of-sample forecasts across 30 countries ($\times 100$ scale) under alternative seasonal adjustment for Southern Hemisphere countries}
    \begin{tabular}{ccccccc}
    \toprule
          & \multicolumn{3}{c}{\textbf{Symmetric seasonal adjustment}} & \multicolumn{3}{c}{\textbf{Reciprocal seasonal adjustment}} \\
    \midrule
    \makebox[1.5cm]{$h$}     & \makebox[1.8cm]{\textbf{LL}} & \makebox[2.1cm]{\textbf{HBY}} & \makebox[2.1cm]{\textbf{GBLL}} & \makebox[1.8cm]{\textbf{LL}} & \makebox[2.1cm]{\textbf{HBY}} & \makebox[2.1cm]{\textbf{GBLL}} \\
    \midrule
    1     & \textbf{5.424} & 6.087 & \textbf{5.262} & 5.436 & \textbf{5.985} & 5.278 \\
    2     & 5.802 & 6.664 & 5.640 & \textbf{5.779} & \textbf{6.512} & \textbf{5.620} \\
    3     & 5.763 & 6.782 & 5.597 & \textbf{5.746} & \textbf{6.537} & \textbf{5.582} \\
    4     & 5.783 & 6.909 & 5.610 & \textbf{5.774} & \textbf{6.547} & \textbf{5.602} \\
    5     & 5.857 & 6.988 & 5.685 & \textbf{5.846} & \textbf{6.553} & \textbf{5.671} \\
    6     & 5.889 & 6.972 & 5.720 & \textbf{5.889} & \textbf{6.563} & \textbf{5.718} \\
    7     & \textbf{5.912} & 6.946 & \textbf{5.745} & 5.923 & \textbf{6.610} & 5.753 \\
    8     & \textbf{5.978} & 6.988 & \textbf{5.811} & 5.998 & \textbf{6.698} & 5.829 \\
    9     & \textbf{6.019} & 7.032 & \textbf{5.852} & 6.048 & \textbf{6.752} & 5.878 \\
    10    & \textbf{6.044} & 7.037 & \textbf{5.875} & 6.085 & \textbf{6.758} & 5.914 \\
    11    & \textbf{6.050} & 6.992 & \textbf{5.877} & 6.096 & \textbf{6.745} & 5.922 \\
    12    & \textbf{6.045} & 6.956 & \textbf{5.866} & 6.091 & \textbf{6.726} & 5.911 \\
    \bottomrule
    \end{tabular}%
  \label{tab:seasonal_transformation}%
\end{table}%

The resulting forecasts in Table \ref{tab:seasonal_transformation} are qualitatively and quantitatively similar to those obtained under the reciprocal transformation. The main conclusion remains unchanged as our proposed GBLL model performs better than the LL model and the HBY model provides the least accurate forecasts. A likely explanation is that there are only two countries in the Southern Hemisphere, namely Australia and New Zealand, and therefore their treatment does not materially affect the overall common trends or seasonal structures identified across the full 30 countries.

\section{Forecast performance evaluated by MASE}
To provide a more comprehensive evaluation of forecast performance, we have included an additional error metric. Given the 30 countries in our sample, which differ in population size and mortality rate scales, we have chosen to include the Mean Absolute Scaled Error (MASE), as it is scale-independent, similar to MAPE. The MASE is defined as the mean absolute error of the forecast, scaled by the mean absolute error of a naive in-sample forecast. Specifically, for $h$-step-ahead forecasts across all age groups, countries, and expanding windows, the MASE is computed using a seasonal naive benchmark with season length $s$ (e.g., $s = 52$ weeks):
\begin{equation*}
    \mathrm{MASE}_h=\frac{1}{4\times H_{(h)} \times 30 \times 10}\sum_{x=1}^4\sum_{u=1}^{H_{(h)}}\sum_{j=1}^{30}\sum_{r=0}^{9}\left(\frac{|\hat{m}^j_{x,169+u+H_{(r)}}-\tilde{m}^j_{x,169+u+H_{(r)}}|}{\sum_{t=52+1}^{208-H_{(r)}}|\tilde{m}^j_{x,t}-\tilde{m}^j_{x,t-52}|/(208-H_{(r)}-52)}\right).
\end{equation*} A MASE value less than 1 indicates that the forecast is, on average, more accurate than the naive benchmark, while a value greater than 1 indicates worse performance.

The results in Table \ref{tab:MASE} lead to conclusions consistent with those based on MAPE. While all three models outperform the naive benchmark, our proposed GBLL model provides more accurate forecasts than both the LL and HBY models, with the HBY model performing the worst. Table \ref{tab:forecasting_MASE_by_age_group} presents a breakdown by age groups, which similarly aligns with the MAPE results.

\begin{table}[h]
    \centering
    \caption{Mean MASE of out-of-sample forecasts across 30 countries}
    \begin{tabular}{cccc}
    \toprule
    \makebox[2cm]{\textbf{$h$}} & \makebox[2.5cm]{\textbf{LL}} & \makebox[3.2cm]{\textbf{HBY}} & \makebox[3.2cm]{\textbf{GBLL}}\\
    \midrule
    1     & 0.640 & 0.732 & \textbf{0.616} \\
    2     & 0.687 & 0.804 & \textbf{0.663} \\
    3     & 0.686 & 0.813 & \textbf{0.661} \\
    4     & 0.692 & 0.819 & \textbf{0.667} \\
    5     & 0.704 & 0.820 & \textbf{0.678} \\
    6     & 0.709 & 0.820 & \textbf{0.685} \\
    7     & 0.715 & 0.827 & \textbf{0.691} \\
    8     & 0.725 & 0.837 & \textbf{0.702} \\
    9     & 0.730 & 0.842 & \textbf{0.707} \\
    10    & 0.734 & 0.841 & \textbf{0.711} \\
    11    & 0.733 & 0.836 & \textbf{0.710} \\
    12    & 0.729 & 0.831 & \textbf{0.706} \\
    \bottomrule
    \end{tabular}%
    \label{tab:MASE}%
    \end{table}%

\begin{table}[htbp]
{\small
  \centering
  \caption{Mean MASE of out-of-sample forecasts across 30 countries by age groups}
    \begin{tabular}{ccccccccccccc}
    \toprule
          & \multicolumn{3}{c}{\textbf{15--64 years old}} & \multicolumn{3}{c}{\textbf{65--74 years old}} & \multicolumn{3}{c}{\textbf{75--84 years old}} & \multicolumn{3}{c}{\textbf{85+ years old}} \\
    \midrule
    $h$     & LL    & HBY   & GBLL  & LL    & HBY   & GBLL  & LL    & HBY   & GBLL  & LL    & HBY   & GBLL \\
    \midrule
    1     & 0.749 & 0.836 & \textbf{0.705} & 0.666 & 0.773 & \textbf{0.640} & 0.596 & 0.726 & \textbf{0.578} & 0.551 & 0.595 & \textbf{0.542} \\
    2     & 0.772 & 0.850 & \textbf{0.730} & 0.709 & 0.850 & \textbf{0.676} & 0.665 & 0.833 & \textbf{0.638} & \textbf{0.601} & 0.684 & 0.608 \\
    3     & 0.765 & 0.845 & \textbf{0.724} & 0.709 & 0.871 & \textbf{0.670} & 0.671 & 0.839 & \textbf{0.640} & \textbf{0.597} & 0.699 & 0.610 \\
    4     & 0.762 & 0.854 & \textbf{0.723} & 0.711 & 0.875 & \textbf{0.670} & 0.687 & 0.827 & \textbf{0.653} & \textbf{0.610} & 0.718 & 0.625 \\
    5     & 0.768 & 0.875 & \textbf{0.729} & 0.715 & 0.865 & \textbf{0.672} & 0.706 & 0.819 & \textbf{0.669} & \textbf{0.625} & 0.724 & 0.644 \\
    6     & 0.771 & 0.887 & \textbf{0.733} & 0.718 & 0.849 & \textbf{0.676} & 0.714 & 0.825 & \textbf{0.675} & \textbf{0.635} & 0.721 & 0.656 \\
    7     & 0.775 & 0.894 & \textbf{0.737} & 0.721 & 0.849 & \textbf{0.680} & 0.720 & 0.841 & \textbf{0.682} & \textbf{0.644} & 0.724 & 0.666 \\
    8     & 0.778 & 0.896 & \textbf{0.741} & 0.728 & 0.855 & \textbf{0.687} & 0.734 & 0.865 & \textbf{0.695} & \textbf{0.660} & 0.732 & 0.683 \\
    9     & 0.779 & 0.898 & \textbf{0.744} & 0.731 & 0.857 & \textbf{0.691} & 0.742 & 0.878 & \textbf{0.702} & \textbf{0.669} & 0.735 & 0.692 \\
    10    & 0.781 & 0.897 & \textbf{0.748} & 0.733 & 0.850 & \textbf{0.694} & 0.747 & 0.882 & \textbf{0.706} & \textbf{0.674} & 0.733 & 0.696 \\
    11    & 0.781 & 0.897 & \textbf{0.750} & 0.731 & 0.839 & \textbf{0.692} & 0.745 & 0.883 & \textbf{0.704} & \textbf{0.673} & 0.727 & 0.693 \\
    12    & 0.782 & 0.898 & \textbf{0.751} & 0.727 & 0.835 & \textbf{0.689} & 0.742 & 0.875 & \textbf{0.699} & \textbf{0.666} & 0.718 & 0.685 \\
    \bottomrule
    \end{tabular}%
  \label{tab:forecasting_MASE_by_age_group}%
  }
    \end{table}%

\section{Supporting figures on fitting and forecast performance}
\subsection{Fitting performance}
To provide a more concrete picture of the in-sample performance, we provide visualisations of the fitted log mortality rates for two representative countries, namely France and South Korea, across all age groups, as shown in Figures \ref{fig:France_fitting} and \ref{fig:SouthKorea_fitting}, respectively. These plots confirm that the LL model captures the general cyclical patterns but fails to fully model the smaller-magnitude fluctuations around these cycles. The HBY model performs the worst, missing both the smaller seasonal patterns and the overall level of mortality rates. In contrast, the proposed GBLL model provides the best fit to the observed data, effectively minimising the discrepancies across both cyclical and smaller-scale variations. 

\setcounter{figure}{0}
\renewcommand{\thefigure}{\thesection.\arabic{figure}}
\renewcommand{\theHfigure}{\thesection.\arabic{figure}}
\begin{figure}[h!]
    \centering
    \includegraphics[width=1\textwidth]{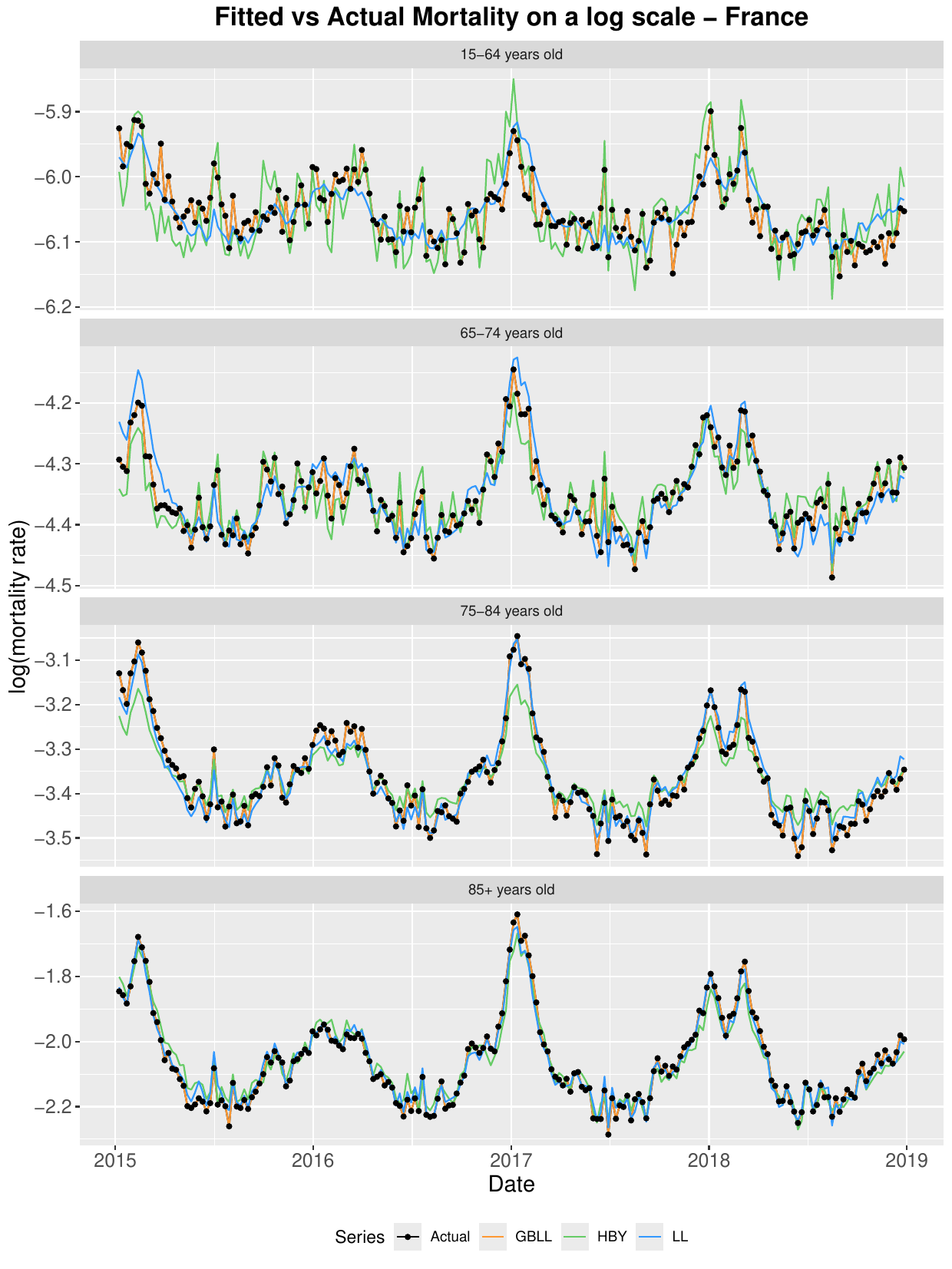}
    \caption{The in-sample fitted mortality rates of France}
    \label{fig:France_fitting}
    \end{figure}

    \begin{figure}[h!]
    \centering
    \includegraphics[width=1\textwidth]{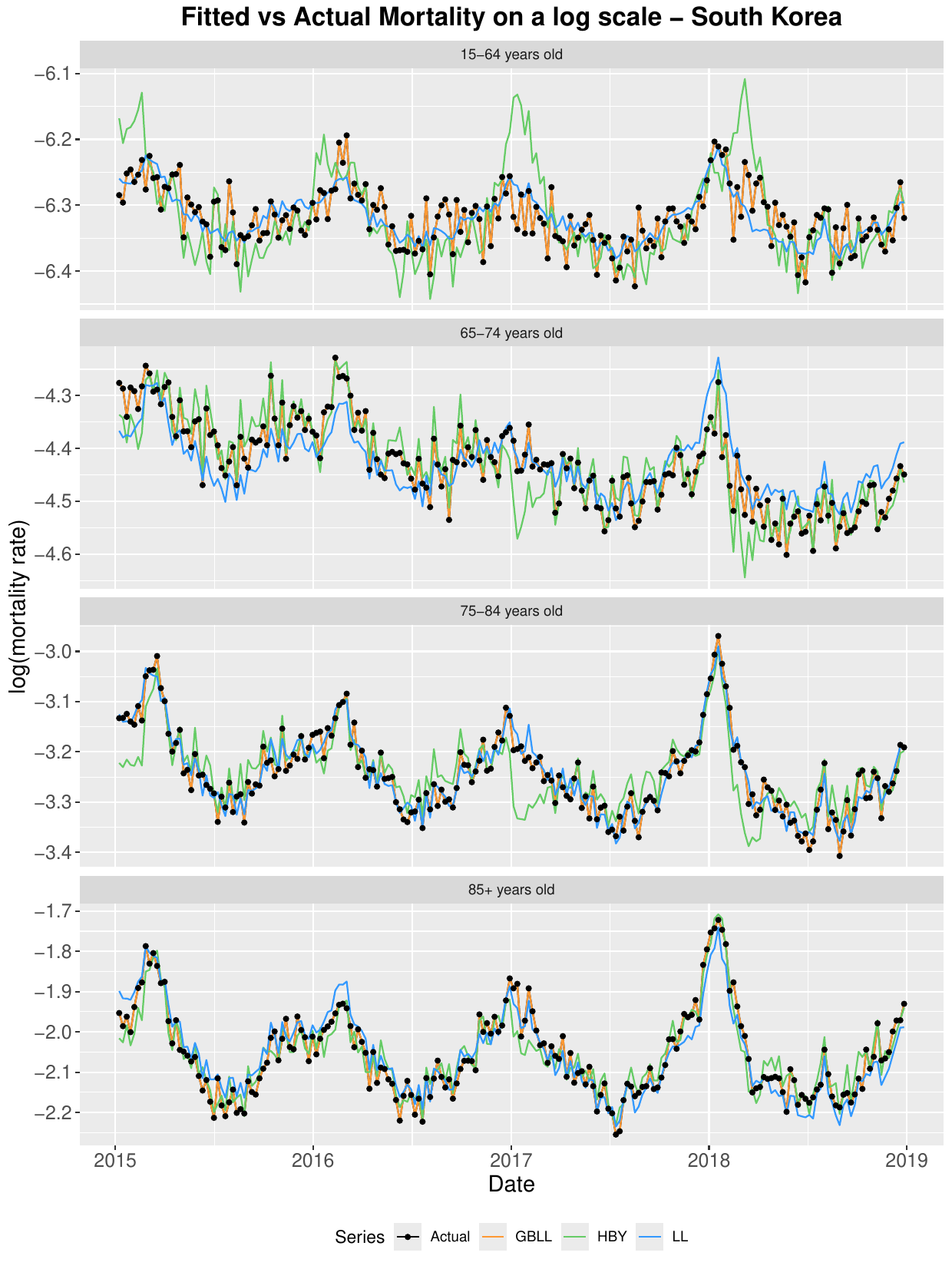}
    \caption{The in-sample fitted mortality rates of South Korea}
    \label{fig:SouthKorea_fitting}
    \end{figure}

\subsection{Forecast performance}
\begin{figure}[h!]
    \centering
    \includegraphics[width=1\textwidth, height=0.37\textheight]{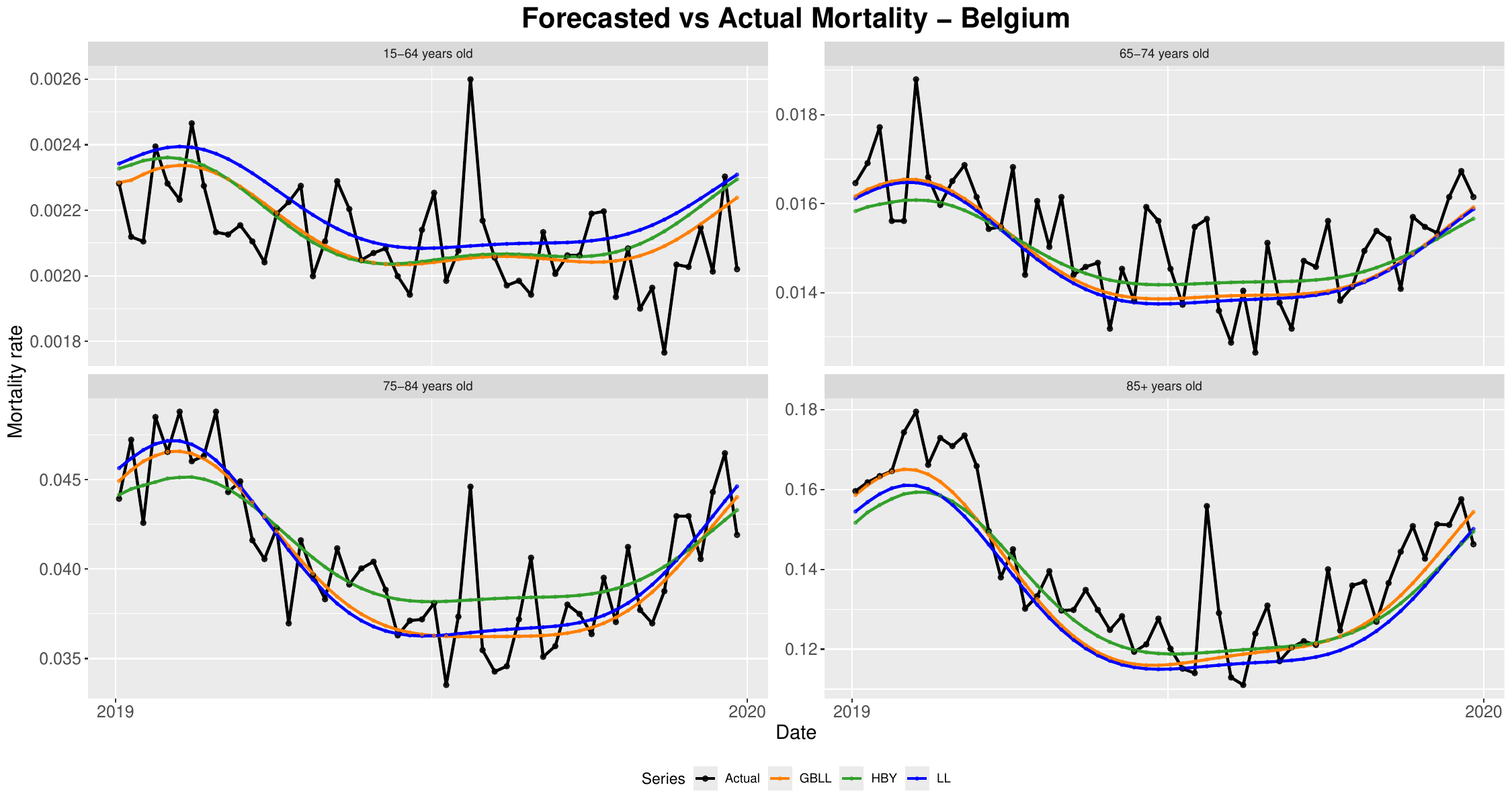}
    \caption{The out-of-sample forecasted mortality rates of Belgium }
    \label{fig:Belgium_forecast}
    \end{figure}

\begin{figure}[h!]
    \centering
    \includegraphics[width=1\textwidth, height=0.37\textheight]{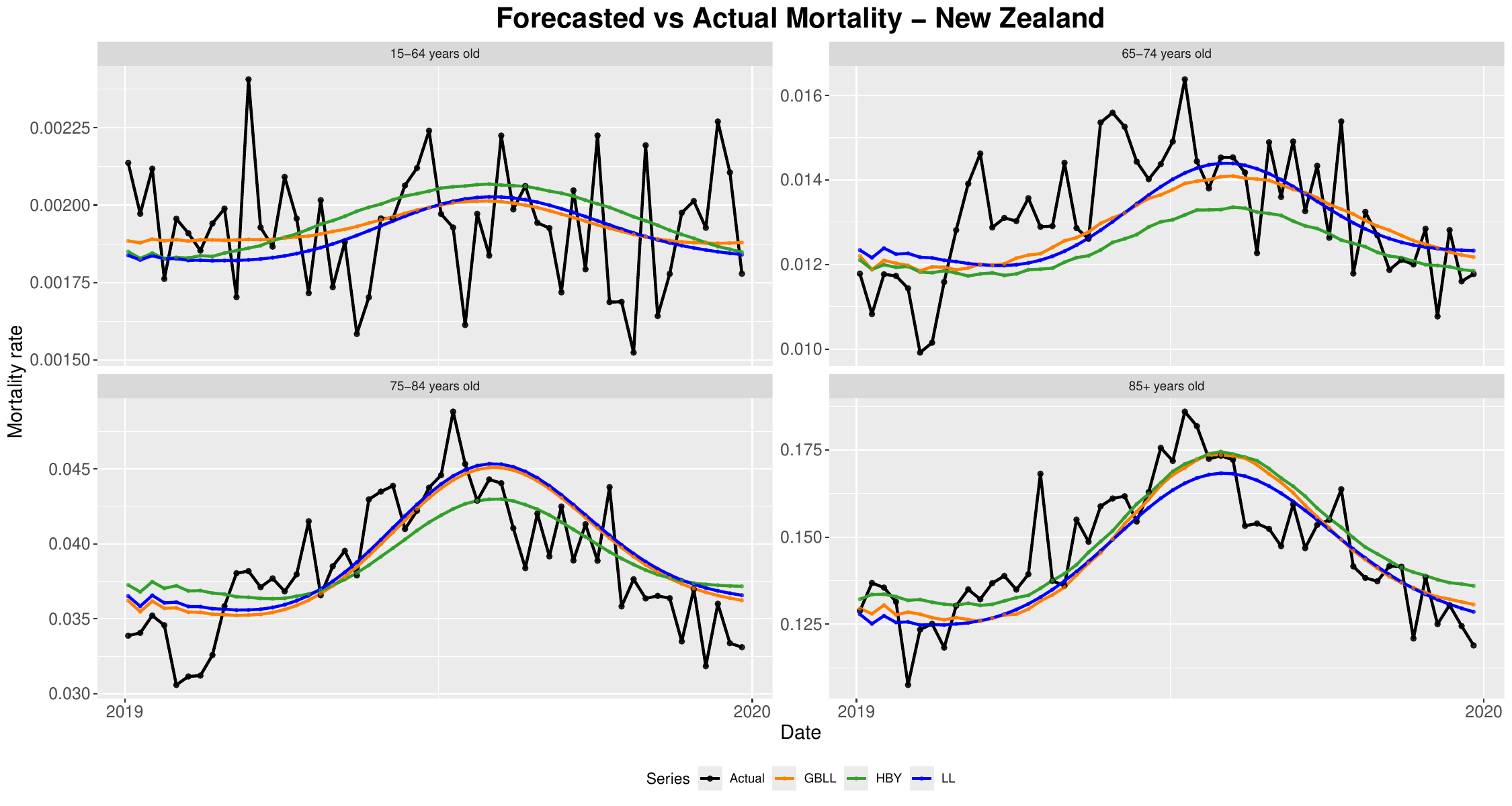}
    \caption{The out-of-sample forecasted mortality rates of New Zealand}
    \label{fig:New_Zealand_forecast}
    \end{figure}

For a visual evaluation of forecast performance, we provide figures comparing the forecasted mortality rates with the observed values for all three models, focusing on Belgium and New Zealand, as shown in Figures \ref{fig:Belgium_forecast} and \ref{fig:New_Zealand_forecast}, respectively. Consistent with the in-sample fitting results, the LL model captures the general seasonal movements but often misestimates the level of mortality. The HBY model, with its six principal components, is more flexible and better captures cyclical movements, but its forecasts tend to deviate from the actual level. The proposed GBLL model provides the most accurate forecasts, correctly capturing both the overall level of mortality and the short-term fluctuations alongside the cyclical patterns, benefiting from the flexibility provided by multiple boosting iterations.

\section{Forecast performance by age groups}
To examine forecast performance across age groups, the results are reported in Table \ref{tab:forecasting_MAPE_by_age_group}. For month $h \in \{1,...,12\}$, expanding window $r \in \{0,...,9\}$, country $j\in\{1,...,30\}$, and age group $x \in \{1,...,4\}$, the age-group-specific MAPE for $h$-step-ahead forecasts, aggregated across all countries and expanding windows, is defined as:
\begin{equation*}
\mathrm{MAPE}_{x,h}=\frac{1}{H_{(h)} \times 30 \times 10}\sum_{u=1}^{H_{(h)}}\sum_{j=1}^{30}\sum_{r=0}^{9}\frac{|\hat{m}^j_{x,169+u+H_{(r)}}-\tilde{m}^j_{x,169+u+H_{(r)}}|}{\tilde{m}^j_{x,169+u+H_{(r)}}}.
\end{equation*}

For the 15--64, 65--74, and 75--84 age groups, the proposed GBLL model consistently delivers the most accurate forecasts, outperforming both the baseline LL and HBY models. Once again, the HBY model exhibits the weakest performance across these groups. However, for the oldest age group (85+ years old), the LL model surpasses the GBLL model for 11 out of the 12 forecast horizons, with the exception of the first month. Nonetheless, the differences in MAPE between the two models are marginal. A possible explanation is that mortality patterns among the 85+ age group exhibit smoother and more dominant seasonal cycles, which are already well captured by the simpler LL model. Consequently, the LL model performs comparably or even slightly better in this case, leaving less room for improvement through the gradient boosting approach. In contrast, the three younger age groups exhibit noisier and less pronounced seasonal behaviours. In this case, the GBLL model, through its iterative process, is better suited for identifying and isolating age-specific dynamics by removing residual cycles that may have been inadvertently imposed by the dominant patterns of the oldest group. As a result, the forecast performance is significantly enhanced.

\begin{table}[htbp]
{\small
  \centering
  \caption{Mean MAPE of out-of-sample forecasts across 30 countries by age groups ($\times 100$ scale)}
    \begin{tabular}{ccccccccccccc}
    \toprule
          & \multicolumn{3}{c}{\textbf{15--64 years old}} & \multicolumn{3}{c}{\textbf{65--74 years old}} & \multicolumn{3}{c}{\textbf{75--84 years old}} & \multicolumn{3}{c}{\textbf{85+ years old}} \\
    \midrule
    $h$     & LL    & HBY   & GBLL  & LL    & HBY   & GBLL  & LL    & HBY   & GBLL  & LL    & HBY   & GBLL \\
    \midrule
    1     & 5.625 & 6.067 & \textbf{5.345} & 5.524 & 6.131 & \textbf{5.354} & 5.166 & 5.996 & \textbf{5.020} & 5.429 & 5.746 & \textbf{5.395} \\
    2     & 5.791 & 6.102 & \textbf{5.501} & 5.783 & 6.676 & \textbf{5.562} & 5.659 & 6.768 & \textbf{5.453} & \textbf{5.883} & 6.503 & 5.964 \\
    3     & 5.754 & 6.040 & \textbf{5.468} & 5.769 & 6.805 & \textbf{5.498} & 5.674 & 6.756 & \textbf{5.439} & \textbf{5.787} & 6.547 & 5.921 \\
    4     & 5.735 & 6.098 & \textbf{5.458} & 5.749 & 6.769 & \textbf{5.458} & 5.760 & 6.658 & \textbf{5.503} & \textbf{5.851} & 6.662 & 5.988 \\
    5     & 5.779 & 6.234 & \textbf{5.501} & 5.769 & 6.687 & \textbf{5.461} & 5.876 & 6.593 & \textbf{5.602} & \textbf{5.959} & 6.699 & 6.122 \\
    6     & 5.810 & 6.314 & \textbf{5.533} & 5.793 & 6.584 & \textbf{5.487} & 5.928 & 6.665 & \textbf{5.645} & \textbf{6.026} & 6.687 & 6.207 \\
    7     & 5.848 & 6.372 & \textbf{5.570} & 5.811 & 6.571 & \textbf{5.506} & 5.955 & 6.802 & \textbf{5.672} & \textbf{6.079} & 6.695 & 6.264 \\
    8     & 5.884 & 6.418 & \textbf{5.610} & 5.861 & 6.611 & \textbf{5.556} & 6.047 & 6.994 & \textbf{5.756} & \textbf{6.199} & 6.770 & 6.392 \\
    9     & 5.909 & 6.454 & \textbf{5.644} & 5.896 & 6.633 & \textbf{5.593} & 6.118 & 7.115 & \textbf{5.817} & \textbf{6.269} & 6.806 & 6.459 \\
    10    & 5.939 & 6.470 & \textbf{5.683} & 5.920 & 6.595 & \textbf{5.619} & 6.165 & 7.162 & \textbf{5.855} & \textbf{6.315} & 6.804 & 6.498 \\
    11    & 5.973 & 6.501 & \textbf{5.721} & 5.917 & 6.534 & \textbf{5.621} & 6.173 & 7.185 & \textbf{5.856} & \textbf{6.322} & 6.760 & 6.490 \\
    12    & 6.011 & 6.543 & \textbf{5.756} & 5.900 & 6.513 & \textbf{5.606} & 6.174 & 7.148 & \textbf{5.845} & \textbf{6.278} & 6.702 & 6.435 \\
    \bottomrule
    \end{tabular}%
  \label{tab:forecasting_MAPE_by_age_group}%
  }
\end{table}%

\end{document}